\documentclass[fleqn,usenatbib]{mnras}
\usepackage{newtxtext,newtxmath}
\usepackage[T1]{fontenc}
\usepackage{ae,aecompl}
\usepackage{graphicx} 
\usepackage{amsmath} 
\usepackage{amssymb} 

\newcommand{\kms}{km\,s$^{-1}$}

\newcommand{\bevii}{$^{7}$Be}
\newcommand{\beviiii}{$^{7}$Be\,{\sc ii}}
\newcommand{\beviiiii}{$^{7}$Be\,{\sc iii}}

\newcommand{\liviiii}{$^{7}$Li\,{\sc ii}}
\newcommand{\livii}{$^{7}$Li}
\newcommand{\liviii}{$^{7}$Li\,{\sc i}}
\newcommand{\nai}{Na\,{\sc i}}
\newcommand{\cai}{Ca\,{\sc i}}
\newcommand{\caii}{Ca\,{\sc ii}}

\newcommand{\iiihe}{$^{3}$He}
\newcommand{\ivhe}{$^{4}$He}

\newcommand{\feii}{Fe\,{\sc ii}}
\newcommand{\fei}{Fe\,{\sc i}}

\newcommand{\crii}{Cr\,{\sc ii}}

\begin{document}
\title[Search for \bevii\ in  novae]{ Search for  \bevii\ in the outburst  of  four recent  novae}
\author[]{ Molaro, P. $^{1,2}$, Izzo, L.$^{3,4}$, Bonifacio, P.$^5$, Hernanz, M. $^6$  Selvelli, P.$^{1}$, della Valle, M. $^7$
\thanks{E-mail: paolo.molaro@inaf.it },
\thanks{Based on  data from Paranal Observatory, ESO, Chile}\\
 $^{1}$  INAF-Osservatorio Astronomico di Trieste, Via G.B. Tiepolo 11, I-34143 Trieste, Italy\\
 $^{2}$  Institute of Fundamental Physics of the Universe, Via Beirut 2, Miramare,   Trieste, Italy\\
 $^3$ Instituto de Astrofisica de Andalucia, Glorieta de la Astronomia s/n, 18008, Granada, Spain\\
  $^4$ DARK Cosmology Centre, Niels Bohr Institutet, K\o{}benhavns Universitet, Lyngbyvej 2, 2100 K\o{}benhavn, Denmark\\
 $^5$ GEPI, Observatoire de Paris, Universit{\'e} PSL, CNRS, Place Jules Janssen, 92195 Meudon, France\\
 $^6$ Institute of Space Sciences (ICE, CSIC) and IEEC, Campus UAB, Cam{\'i} de Can Magrans s/n, 08193 Cerdanyola del Valles (Barcelona), Spain\\
 $^7$ Capodimonte Astronomical Observatory, INAF-Napoli, Salita Moiariello 16, 80131-Napoli, Italy}
\date{Accepted.... Received 2018...}
\pagerange{\pageref{firstpage}--\pageref{lastpage}} \pubyear{2002}
\maketitle
\label{firstpage}
\begin{abstract}
Following the recent detection  of \beviiii ~  in the   outburst spectra of  Classical Novae  we  report  the search for this  isotope in the outbursts of four recent  bright novae  by means of high resolution   UVES  observations.   The \beviiii ~  $\lambda\lambda$313.0583,\,313.1228 nm  doublet  resonance lines    are   detected in  the high velocity components of Nova  Mus 2018 and ASASSN-18fv   during outburst. On the other hand \beviiii\ is neither  detected in  ASASSN-17hx and  possibly  nor in Nova Cir 2018, therefore  showing that the \bevii\  is not always ejected in the thermonuclear runaway. Taking into account the  \bevii~ decay we find   $X(\mbox{\bevii})/X(\mbox{H})$   $\approx$ 1.5 $\cdot 10 ^{-5}$ and 2.2 $\cdot 10 ^{-5}$ in Nova Mus 2018 and ASASSN-18fv, respectively. A value of \bevii/H $\approx$  2 $\cdot 10 ^{-5}$  is found in 5 out of the 7  extant measurements   and it might be considered as a typical \bevii\ yield for novae. However, this value  is almost one order of magnitude larger than predicted by current theoretical models.  We argue  that  the variety of high \bevii/H abundances  could be originated  in a higher than solar   content of \iiihe\ in the donor star. The cases with  \bevii\ not detected might be related to a small mass of the WD or to relatively little mixing with the core material of the WD. The    \bevii /H, or  \livii/H,  abundance  is  $\approx$ 4 dex  above   meteoritic  thus  confirming   the novae as  the main  sources of \livii~  in the Milky Way. 
\end{abstract}
\section{Introduction}
In the Classical Nova model hydrogen rich material is transferred  from a main-sequence star or evolved giant through an accretion disk onto a white dwarf. The temperature onto the WD surface rises until CNO cycle fusion starts \citep{Starrfield2016}.    It was suggested that in the
thermonuclear runaway also \bevii\ could  be made via the reaction \iiihe($\alpha,\gamma$)\bevii\ and then ejected   \citep{Arnould1975,Starrfield1978}. 
\bevii\ is a radioactive nucleus and decays into \livii\  with a half-life of 53.22 days so that all \bevii\ is expected to decay into \livii. 
This  suggestion  dates back  in  the 70's  but thwarted by the non detection of \livii\ \citep{Friedjung1979}. 
The long sought \liviii\  $\lambda\lambda$ 670.8 nm resonance line  was recently identified  in the  slow nova  V1369 Cen  \citep{Izzo2015}. On the other hand,   the resonance doublet of the parent nucleus  $^7$Be II   was detected in  few   novae   \citep{Tajitsu2015,Tajitsu2016,Molaro2016,Izzo2018}. By re-analysis of  historical novae by means of archival IUE observations
  \beviiii\ has been  detected also in emission  in the very fast  nova   V838 Her  \citep{Selvelli2018}. This detection  shows  that \bevii\   is  freshly created
 in the nova thermonuclear runaway and ejected in the outburst. It must be noted that in all these cases  \liviii\ at   $\lambda\lambda$  670.8 nm went undetected. So far   \liviii\  has been detected only  in V1369 Cen  while the parent nuclei    \bevii\   has been    detected in all   the  novae where it has been searched for. The  persistent non-detection of neutral \livii\  after more than a decaying time  has been explained with  a capture of a K-shell electron by  \bevii~  thus transforming into  \liviiii~ which    does not neutralize  throughout the time of the outburst \citep{Molaro2016}.   
 
 The \beviiii\ $\lambda\lambda$ 313.1 nm  doublet shows  a huge equivalent width  comparable only to hydrogen and  much greater  than  all other elements.  The measured atomic fraction implies   massive \bevii\ ejecta with the final \livii~ product  up to 4 and even  5 orders of magnitude above meteoritic   abundance. With such an overproduction novae could be  an important   $^7$Li factory   exceeding by far what    synthesised in the  Big Bang, or by other sources such as spallation by cosmic rays and AGB stars.  The role of novae as   \livii~ producers has been considered by means of a detailed model of the
  chemical evolution of the Milky Way   \citep{Cescutti2019}. They    showed that novae  account well for the observed
  increase of Li abundance with metallicity  in the Galactic thin disk and also for the relative flatness observed  in the  thick disk. The latter evolves on a timescale which is  shorter than the typical timescale in which  novae   produce substantial \livii\ and therefore do not show any enhancement.

A major problem  is that the abundances measured exceed what foreseen by  current models  \citep{Hernanz1996, Jose1998}. 
Initially, \bevii\  detection occurred  in slow novae, but  later  it  has been found also in   fast novae such as  V407 Lup, whose progenitor is likely a ONe WD \citep{Izzo2018},  and  V838 Her which is one of the fastest ever observed \citep{Selvelli2018}. Thus, it seems that \bevii\ is present in both fast and slow novae   and with comparable abundances.  
  
The decay  of \bevii\  produce a  high-energy line at 478 keV emitted during the de-excitation to the ground state of the fresh \livii\  produced in the \bevii\  electron-capture \citep{Gomez1998}. Several unsuccessful attempts to detect the line with Gamma ray satellites have been performed \citep{Harris2001}. The  detection of the radioactive \bevii\ nuclei  in the nova outburst reopened the possibility of detectability  of the 478 Kev line with INTEGRAL for  nearby novae. The   distance should be less then $\approx$ 0.5 Kpc, though the horizon will depend on the amount of \bevii\ produced in the nova event \citep{Siegert2018}. 
The nova ASASSN-18fv was observed with INTEGRAL-Director Discretionary Time, during 2.8 Ms. Although the complete data analysis is still ongoing, it is already clear that only upper limits to the 478 Kev emission line have been obtained, which are not constraining for the models (Siegert et al in prep).
    
  Following the recent detection  of \beviiii ~  in the   outburst spectra of  Classical Novae,  we  activated a ToO program at ESO to target \bevii\ in  novae which at maximum reach magnitude  V $\le$ 9. We report here on the search for the \bevii~ isotope in  high resolution   UVES  spectra  of four recent  novae  discovered in the years 2017 and 2018.


\section{The program novae}

 The four novae of this program have been    the  brightest ones  since 2017.  Source brightness is  required  by the need to study  the  resonance doublet lines \beviiii ~  which lay at $\lambda\lambda$313.1 nm,   very close to the atmospheric cutoff  where  a significant atmospheric absorption is present.

Several UVES spectra   for each nova  were obtained following the   maximum.  The UVES settings were DIC1 346-564,  with central  wavelength of 346\,nm (range 305-388\,nm) in the blue arm, to cover the  \beviiii~ $\lambda\lambda$313.1 nm  lines and  564\,nm (460-665\,nm) in the red arm. Every observation was followed by another with setting  DIC2 437-760,  
to cover in the blue the H \& K  CaII lines  and   in the red arm to cover several  metallic lines and the  \liviii\  at $\lambda\lambda$ 670.8 nm.    The journal of the observations for each nova is  provided in  Table \ref{tab:1}. The resolving power   was typically $R= \lambda /\delta \lambda  \approx 80,000$ for the blue arm and $\approx 120,000$ for the red arm.  Overlapping spectra have been combined for each epoch to maximise the signal-to-noise ratio.

\begin{table}\label{tab:1}
\caption{Journal of the observations. Only  DIC1 settings are reported. In correspondence of each DIC1 several DIC2 setting observations were taken with  high resolution. }
\begin{center}
\scriptsize
\begin{tabular}{rrrrrr}
\hline
\hline
\multicolumn{1}{c}{{MJD}} & 
\multicolumn{1}{c}{{Day}} & 
\multicolumn{1}{c}{exp (s)}& 
\multicolumn{1}{c}{R/1000}& 
\multicolumn{1}{c}{exp (s)}&
\multicolumn{1}{c}{R/1000}\\
\multicolumn{1}{c}{} & 
\multicolumn{1}{c}{a.m.}&
\multicolumn{1}{c}{346 nm} &
\multicolumn{1}{c}{} & 
\multicolumn{1}{c}{564 nm} &
\multicolumn{1}{c}{}  \\
\hline
\multicolumn{1}{c}{ASASSN-17hx}\\
\hline
   57980.0830&  15 &  1100 &  71 &1100  & 107 \\
   57988.0978&  23 &  1100  & 50 &1100  & 52\\
    58013.0586&  48 &  1100 & 50     &  1100 &52\\
    58056.0242 & 91  & 1100 & 71 &1100& 107  \\
   \hline
   \multicolumn{1}{c}{NOVA MUS 2018}\\
   \hline
  58134.3246  & 0  & 1200 &50   &&  \\
  58134.3246  &  & 300&50   && \\
   58169.0644  & 35  & 800 & 52   &800 &52 \\
   58169.0877 &   & 600&50   & 600& 52  \\
   58169.0952 &   & 600&50    & 600&52 \\
   58174.1406& 40  & 1200&50  &  1200 & 52 \\
   58178.1866& 44  & 1200&50  &  1200& 52 \\
    \hline
   \multicolumn{1}{c}{NOVA CIR 2018}\\
   \hline
  
   58142.3528& -15  & 60 & 71     & 60 &107   \\
   58142.3542&   & 300 & 71  &  300 & 107   \\
   58150.2844& -7  & 600 & 59     & 600& 66  \\
    58169.2516& 12  & 600 &50     & 600& 52  \\
     58169.2598&   & 200 & 50     & 200& 52  \\
      58169.2627&   & 200 &50     & 200&52  \\
       58169.2656&   & 200 & 50     & 200 &52  \\
       58177.1187& 20   & 600& 50    & 2x250 & 52\\
       58181.3439& 24   & 600  & 50  & 2x250 & 52 \\
       58187.1453& 30  & 600 & 50  &   2x250 & 52 \\
        58191.3409& 34  & 600 & 50    & 250 & 52 \\
        58215.1463& 58  &  600 & 59  &   120 & 66   \\
        58220.1456& 63  &  600 & 59     & 120 & 66   \\
         58223.1423& 66  &  600 & 59     & 60 & 66   \\
        58227.9869& 71  &  300 & 59     & 60 & 66   \\
        58232.2215& 75  &  300 & 59     & 120 & 66   \\
        58232.2136&   &  600 & 59     & 120& 66   \\
         58232.2054&   &  600 & 59     & 120 & 66   \\
           58243.0936& 86  &  600 & 59     & 60 & 66   \\
        \hline
   \multicolumn{1}{c}{ASASSN-18fv}\\
   \hline
    58199.1438& -7  & 60 & 59    & 60&66   \\
    58199.1452&   & 300&59     & 300&66   \\
    58201.1412& -5  & 300&59     & 60&66   \\
    58203.0924& -3  & 300&59     & 60&66   \\
    58205.0687& -1  & 300&59     & 60&66   \\
    58207.0599& 1  & 600&59     & 120&66   \\
    58209.0218& 3  & 600&59     & 120&66   \\
    58213.0665& 7  &  600&59     & 120&66   \\
    58215.1292& 9  &  600&59     & 120&66   \\
    58217.0686& 11  &  600&59     & 120&66   \\
    58220.1292& 14  &  600&59     & 120&66   \\
    58223.1262& 17  &  600&59     & 60&66   \\
    58227.9869& 21  &  300&59     & 15&66   \\
    58228.0002& 22  & 150&59    & 30&66   \\
    58228.0043&   &  50&59     & 6&66   \\
    58235.9959& 29  &  300&59     & 60&66   \\
    58249.0205& 43  &  600&59     & 120&66   \\
    58262.1078& 56  &  600&59     & 120&66   \\
    58280.0381& 74  &  1200&59     & 3x224&66   \\
    58285.9707& 80  &  1200&59     & 120&66  \\
    58303.9851& 98  & 1400&59     & 13x60&66   \\
        \hline

\hline
\end{tabular}
\end{center}

\end{table}

The \beviiii\ resonance lines are  at 313.1  nm  in a  particularly challenging region  due  both  to    earth's atmosphere absorption   and by transmission optical components   of the spectrographs.    The identification was made   by the close correspondence  in the position of the \beviiii~ lines with that of other  ions which are known to be present in the outburst spectra. 
Coincidences are always possible and the most   strong lines in the \bevii\ region are listed  in  Table \ref{tab:2}. 
The detection becomes robust  only  when also  the other line of the \bevii\ resonance doublet at 313.1228 nm is   visible.  Unfortunately,   narrow components are not always present  in the outburst spectra. When   seen the UVES resolution    of $\approx$ 4 \kms allows to distinguish even between  \bevii~ and $^9$Be isotopes. $^9$Be transitions fall at  $\lambda\lambda$ 313.04219 and 313.10667 nm   at    15.6 \kms  from the lighter isotope. However, $^9$Be is not synthesized in the Nova thermonuclear runaway and if present it would be the one from the companion at a  level of $^9$Be/H= $2.4 \times 10^{-11}$  \citep{Lodders2009}. Such a small abundance would be undetectable in the  outburst spectra and  \bevii~ is found  at  an abundance of  six orders of magnitude higher.

\subsection {Nova ASASSN-17hx}
ASASSN-17hx    was  discovered on June 19,  2017 \citep{Stanek2017a,Stanek2017b,Stanek2017c} in the Scutum constellation by the All Sky Automated Survey for SuperNovae (ASASSN) survey \citep{Shappee2014}. Initially of V=12 mag, ASASSN-17hx  brightened continuously  to reach  V $\approx$  8 mag  on 31 July 2017.  
The nova showed  a long  pre-maximum plateau. This is  characteristic of  slow novae and   produced by an  
 optically thick expanding envelope that undergoes cooling during the expansion.  
After peak  the nova declined and re-brightened  several times     up to one hundred days after maximum. 
 The detailed  light curve of  ASASSN-17hx is shown in Figure \ref{fig:1a}.
The nova was  initially classified as a He/N nova,  but later developed   strong \feii\ features, thus showing  features of both He/N and Fe II novae at different epochs \citep{Guarro2017,Munari2017,Pavana2017,Williams2017,Poggiani2018}.
 ASASSN-17hx is at low Galactic latitude ($b= -2^\circ.22442$)  and  it is highly reddened.  The reddening maps of \citet{SF11}
provide $E(B-V)= 1.565$ mag while those of \citet{Schlegel} provide $E(B-V)= 1.820$ mag, the
latter value becomes $E(B-V)= 1.218$ mag, when corrected according to the prescription
of \citet{BMB}. Our spectra provide an independent estimate of the reddening from
the diffuse interstellar band (DIB) at 578.0\,nm using the relation of \citet{dib}.
We measure an equivalent width of the DIB of 0.05797\,nm that implies $E(B-V)= 1.139$ mag.
Adopting this value and $E(G_{BP}-G_{RP}) =  0.41595 A_V$ \footnote{See \url{http://stev.oapd.inaf.it/cgi-bin/cmd}, this value is derived using the \citet{ODonnell} extinction curve. } 
we derive a colour for the likely nova progenitor $(G_{BP}-G_{RP})= -0.202$ mag, that corresponds to a black-body with a temperature of about 17\,000\,K.

\begin{figure}
\includegraphics[width=0.99\columnwidth]{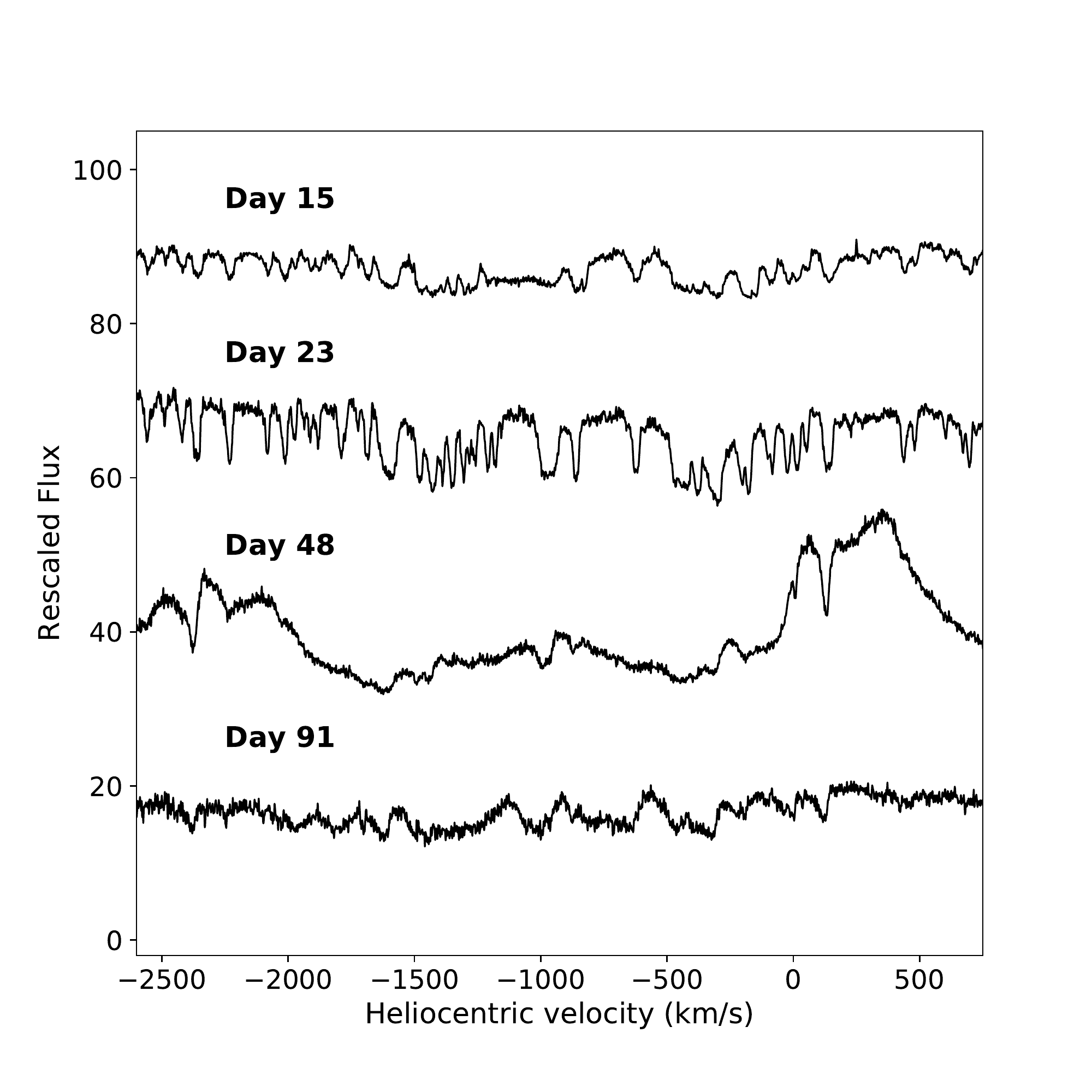}
\caption{ Evolution of ASASSN-17hx in the region of  \bevii. Spectra are in heliocentric velocities with  the zero of the scale  set at \beviiii $\lambda 313.0583$ nm. }\label{fig:2a}
\end{figure}

UVES spectra of ASASSN-17hx have been taken at four epochs starting from day 15 after the maximum at JD 2457964.4886  until  day 91. The region around the \bevii\ at these epochs is shown in Figure \ref{fig:2a}.  The spectra of day 15 and day 23 are  very similar but quite different form the other epochs.  In the early spectra   common features are  strong absorptions    with relatively low velocity at $\approx$ -300 \kms and  at $\approx$ -450 \kms and a broad high velocity component spanning 200 \kms  with a mean velocity at $\approx$ -1000 \kms. 

\begin{figure*}
\includegraphics[width=0.99\columnwidth]{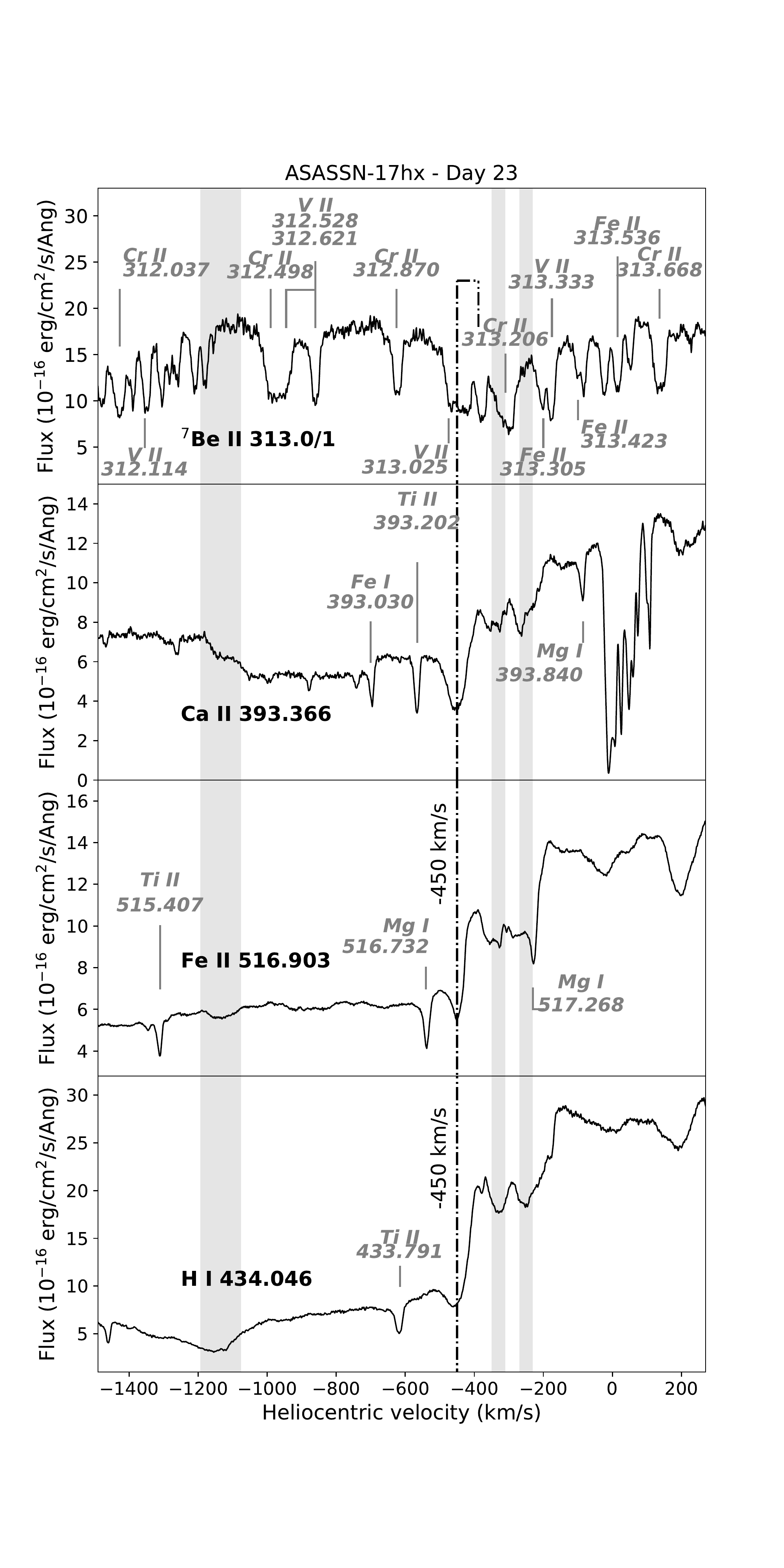}
\includegraphics[width=0.99\columnwidth]{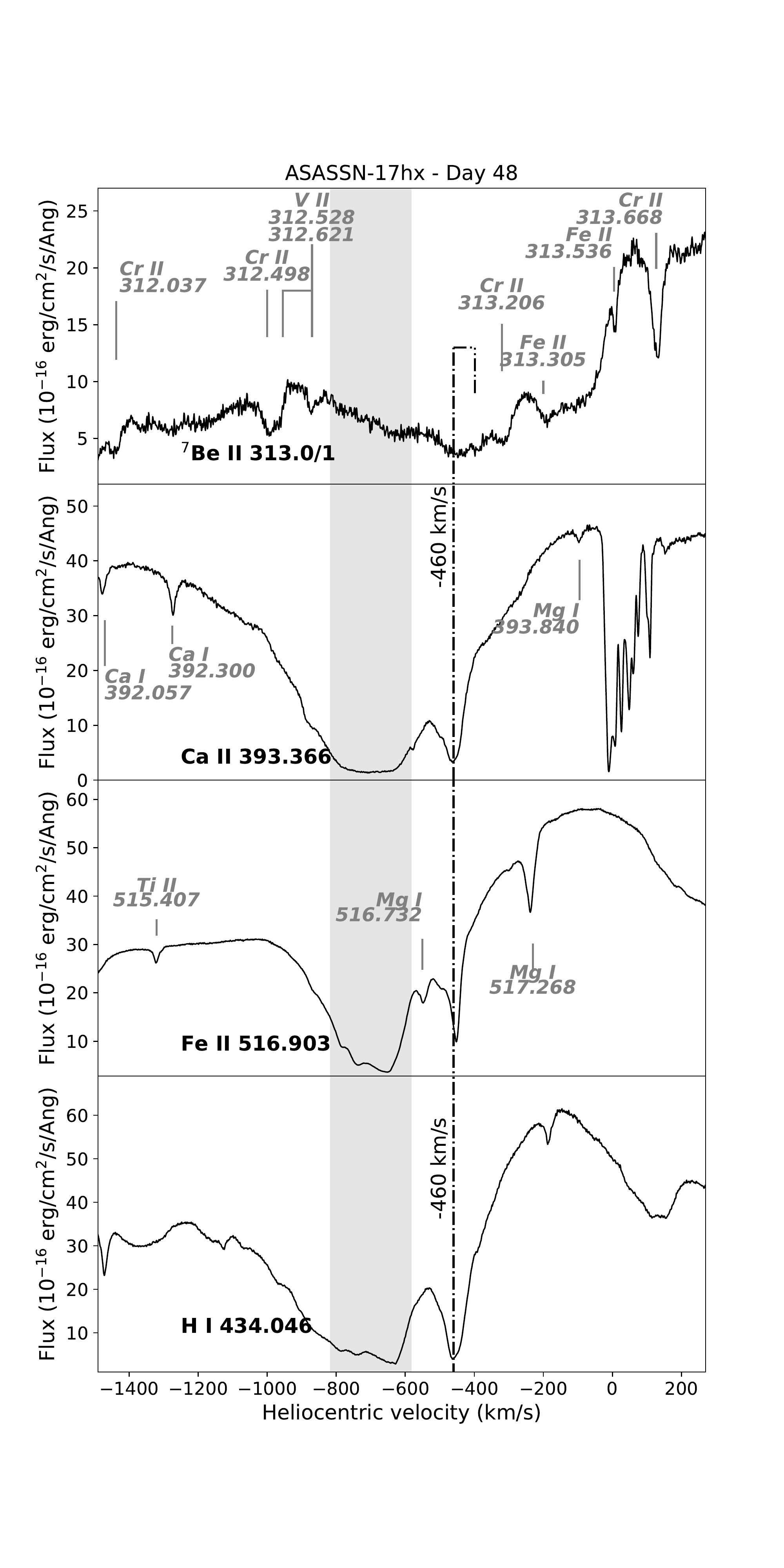}
\caption{ Left: spectrum  of  ASASSN-17hx of day 23 in the region of  \beviiii\  and compared with the  \caii\ K,  \feii ~$\lambda$ 516.903 nm and H$\gamma$  lines. All   identifications shown in  in the panels refer to the -450 \kms component. At zero  and slightly positive  velocities there are multiple    components due to the interstellar  \caii\ K-line. Right: the spectrum of the nova at  day 48.  Note that at this epoch the main absorption component  is found  at  -460 \kms. X-axis  as  in Figure \ref{fig:2a}.  \label{fig:2b}}
\end{figure*}

The spectrum of day 23  is shown in  the left panel of   Figure \ref{fig:2b} where  \beviiii,   \caii\ K,   \feii $\lambda$ 516.90 nm  and  H$\gamma$    regions  are plotted onto a common velocity space.   A close  correspondence  could be be found only for the -450 \kms component as marked in the figure. However, the correspondence is lost for the other absorptions and we conclude that \beviiii\ is probably not present in the nova outburst spectrum.  The  spectra of later epochs look  quite different showing a marked evolution of the outburst.  The spectrum at day 48  is shown in the right panel of Figure \ref{fig:2b}. In  addition to   the -460 \kms component shows the presence of a new and very strong component at  -700 \kms,   while the very high velocity component   vanished away.
In the last epoch  the new component decreased considerably, although remaining quite strong, and moved slightly to  $\approx$ -800 \kms. 
In both these  epochs no correspondence is found between \caii\ and a possible presence of \bevii\   and  we conclude that there is no evidence for the presence of \beviiii\ in the spectrum. 
Due to the general weakening of the absorption 
  there are several absorption features which  can be identified in the \bevii\ regions as    \crii\ lines of 313.668, 313.206 and 312.870 nm. This supports the conclusion that \beviiii\ is  either  weak or  totally absent. Of the    few  investigated so far  this is the first nova for which it is possible to rule out  with some confidence  the presence of \beviiii\ in the outburst spectra. 
The failure of  detection of $^7$Be  should  hold important bearings on the thermonuclear mechanisms leading to the \bevii\ synthesis and ejection.

\subsection{Nova Mus 2018}
\begin{figure}
\includegraphics[width=1.0\columnwidth]{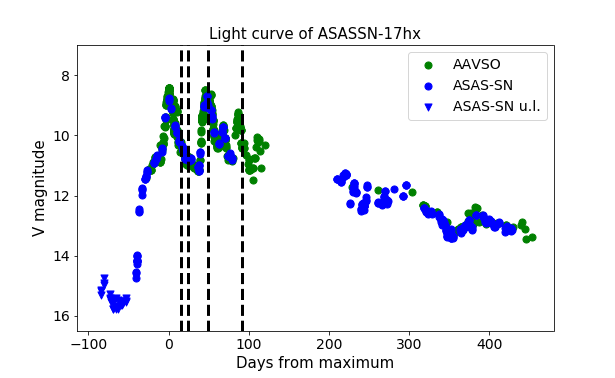}
\includegraphics[width=1.0\columnwidth]{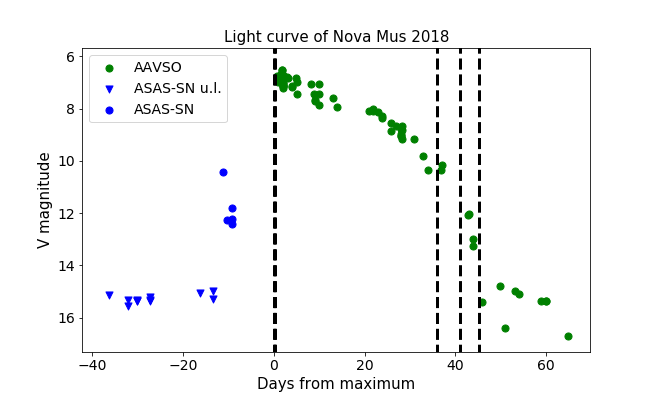}
\includegraphics[width=1.0\columnwidth]{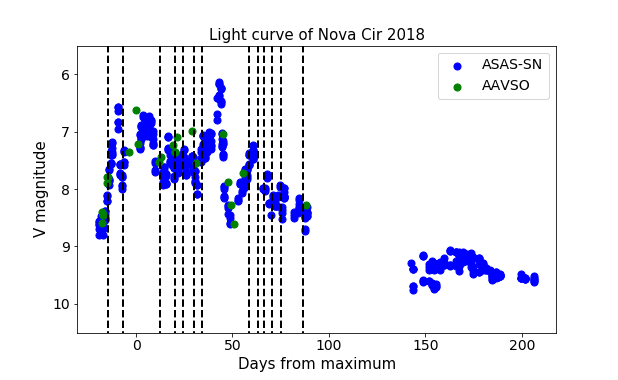}
\includegraphics[width=1.0\columnwidth]{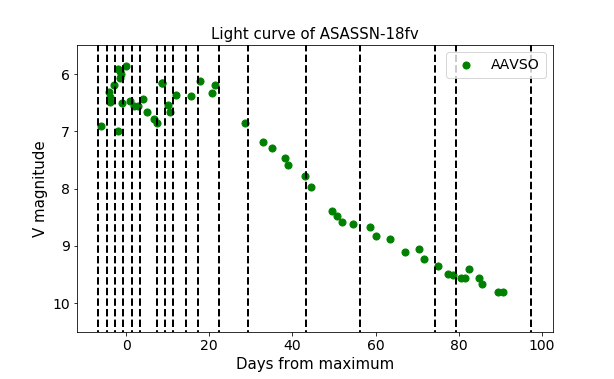}
\caption{V-band light Curves of the four novae described in this paper, obtained using data from the ASASSN survey and the AAVSO database. From the top, ASASSN-17hx, Nova Mus 2018, Nova Cir 2018 and ASASSN-18fv. Vertical lines mark our spectroscopic observations. The maximum epoch for each nova has been computed from AAVSO data. These are: JD 2457964.4886 for ASASSN-17hx; JD 2458134.8891 for  Nova Mus 2018;JD   2458157.0024 for 
Nova Cir 2018       and JD     2458205.9593 for 
ASASSN-18fv}\label{fig:1a}
\end{figure}

 Nova Mus 2018, also PNV J11261220-6531086, was discovered on January 14, 2018 by  Rob Kaufman (2018,  AAVSO Alert Notice 609) in the Musca constellation with a visual magnitude of $\approx$ 7.0 mag and then confirmed as  a Fe II nova. Archival data of ASASSN Sky Patrol observations    showed  that the nova eruption began approximately  two weeks earlier (P. Schmeer, AAVSO Alert Notice 609). The light curve of Nova Mus 2018 is shown in Figure \ref{fig:1a}.
 
\begin{figure}
\includegraphics[width=0.99\columnwidth]{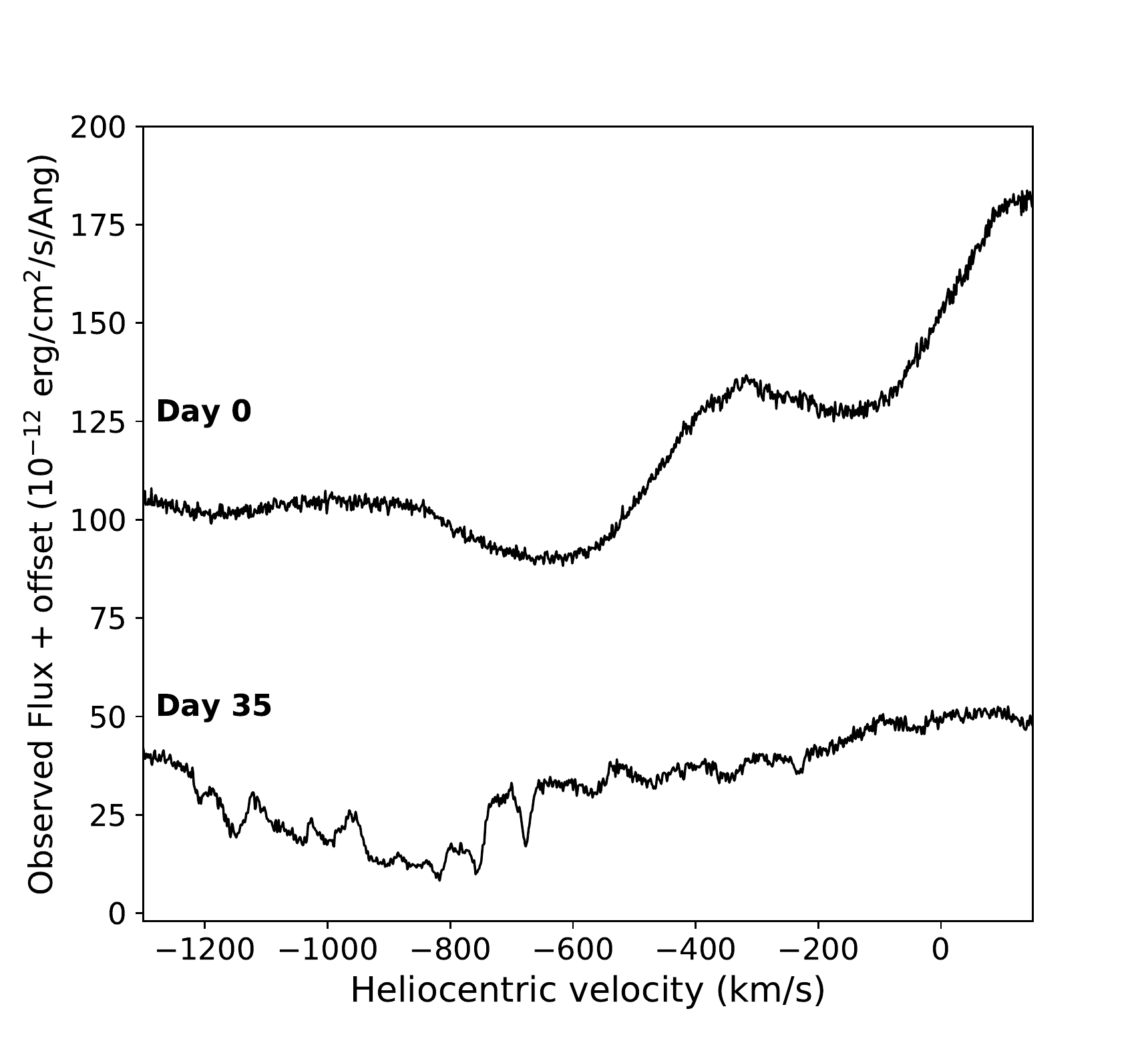}
\caption{   Evolution of Nova Mus 2018 in the region of \bevii. X-axis  as  in Figure \ref{fig:2a}.}\label{fig:3a}
\end{figure}

\begin{figure}
\includegraphics[width=1.0\columnwidth]{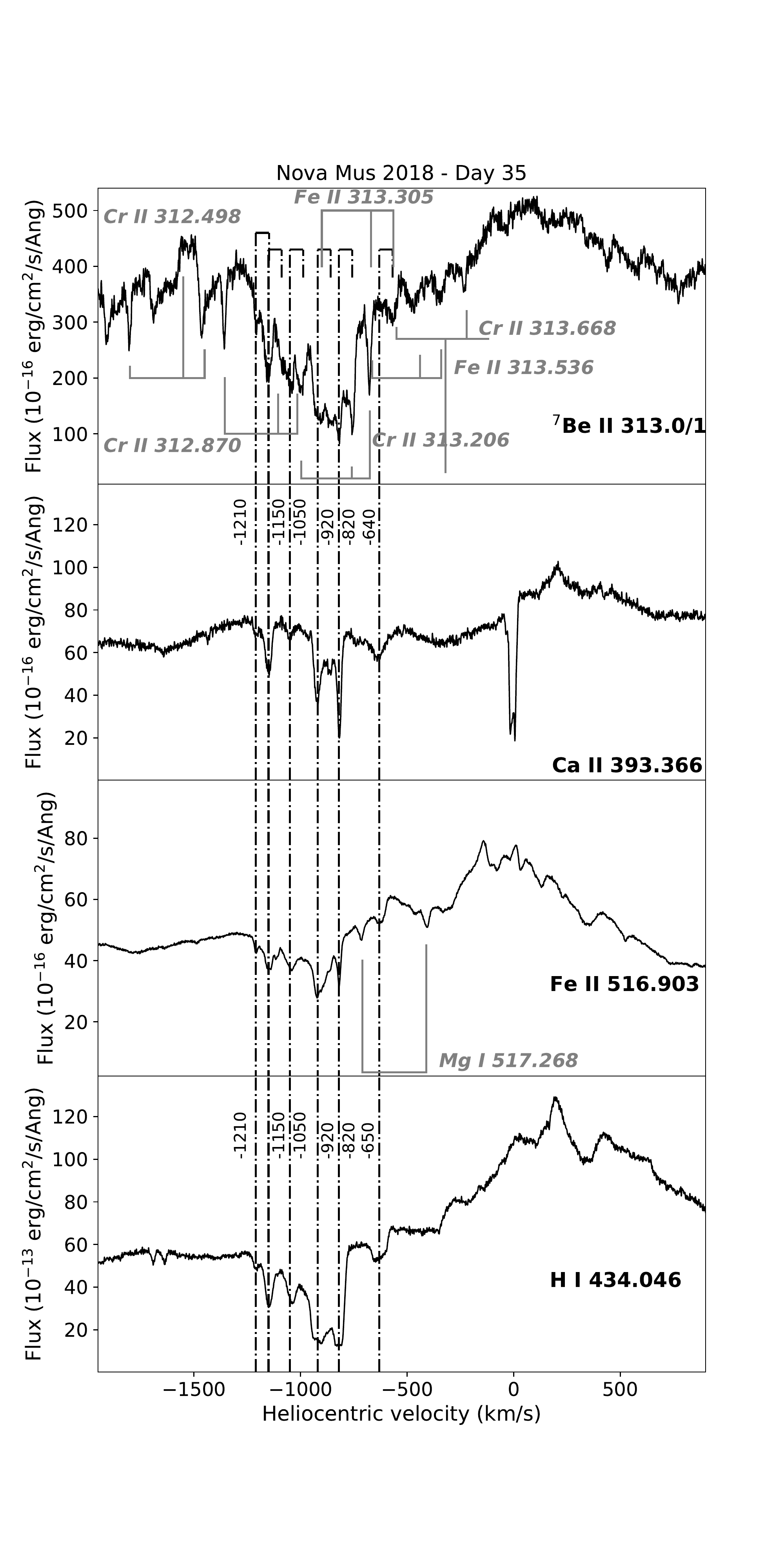}
\caption{  Spectrum of Nova Mus 2018 at  day 35. The blends identified in the \beviiii\ panel refer to the three main components with velocities of -820, -920 and -1150 \kms, respectively. X-axis  as  in Figure  \ref{fig:2a}.}\label{fig:3b}
\end{figure}

\begin{figure}
\includegraphics[width=1.0\columnwidth]{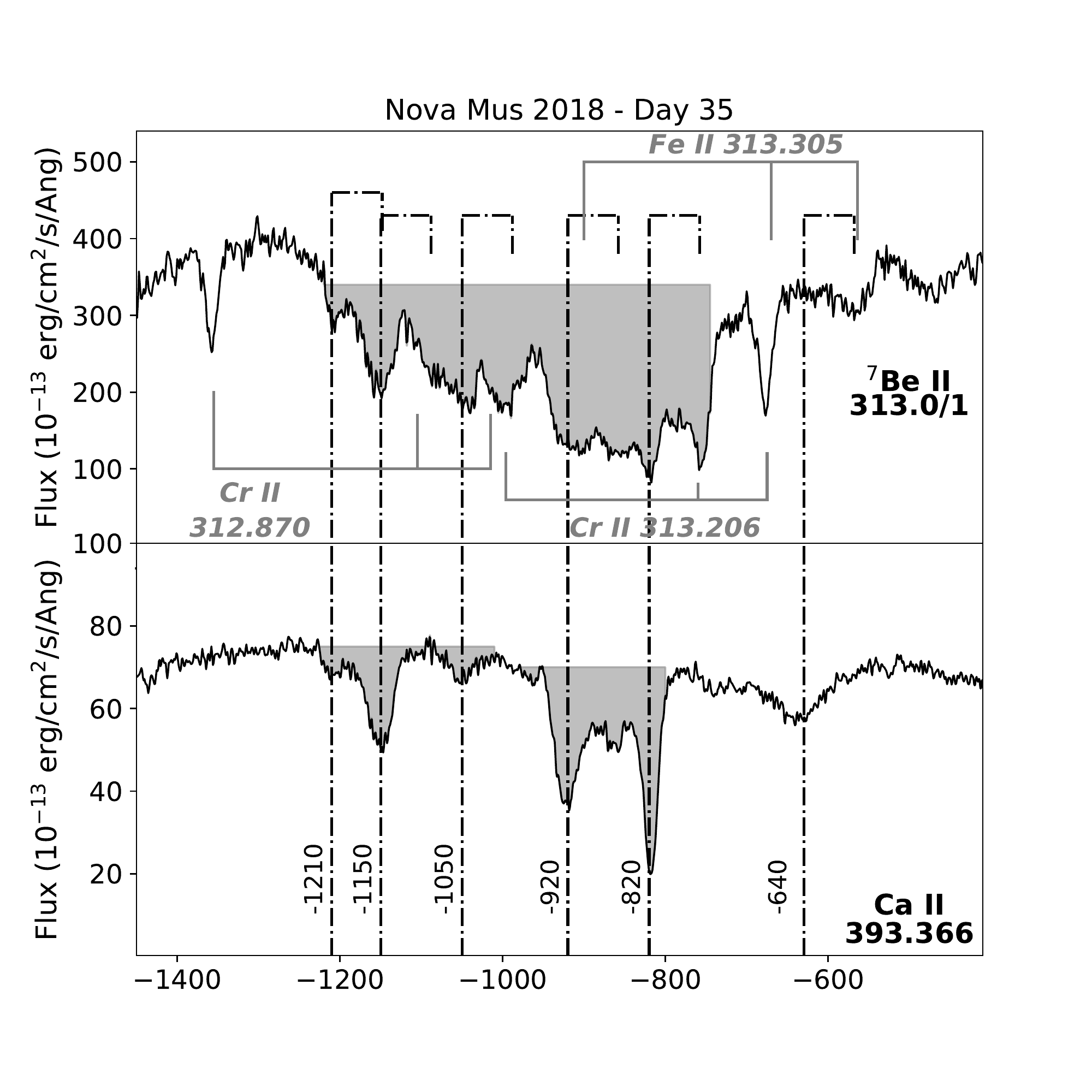}
\caption{  Magnification of   the spectrum of Nova Mus 2018 at  day 35 to show the \beviiii\ identifications.}\label{fig:3c}
\end{figure}

Nova Mus 2018 faded quite rapidly and we could take four epochs which include one epoch on the maximum brightness and other three covering  up to 45 days after maximum when the nova was already of 14 mag, just before the beginning of the observed dust formation dip. The nova  became a very difficult target for observations at  \beviiii $\lambda$ 313.1 nm  and the spectra at day 40 and day 45 are severely underexposed in the  region.  The early spectra are shown in Figure \ref{fig:3a}. The spectrum at day  35  in the \beviiii\    region is  compared with the  \caii\ K,  \feii $\lambda$ 516.903 nm and H$\gamma$  lines  in Figure \ref{fig:3b}. The spectrum   shows a relatively simple structure with three  main absorption components at  -820, -920 and -1150 \kms and a number of smaller ones. All these components  have  a clear correspondence with the stronger of  the  \beviiii~ 313.1 doublet   and show often also  evidence for  the fainter  one.   A zoom of   the spectrum of Nova Mus 2018 at  day 35 is shown in Figure \ref{fig:3c}. Thus, there are  no doubts about    the identification and we consider it as a robust   detection of the \beviiii.   

\begin{table}
\caption{A list of single-ionized ions which are the main   contributors   to
the absorption in the region around  $\lambda$313.0~nm.  Atomic data  taken from the  NIST lines data base. Those for \crii\ are from \citet{lawler2017L}. }
\label{tab:2}
\begin{center}
\begin{tabular}{crccc}
\hline
\hline
Wavelength (Air) & ion   & Log(gf) & Low en. & Upper en.  \\
nm &  &  &   (eV) & (eV) \\
\hline
\hline
   311.0672 & Ti {\sc ii} &        -0.953   & 1.23 & 5.21 \\
   311.2049 & Ti {\sc ii} &        -1.157   & 1.22 & 5.21 \\
   311.4295 & Fe {\sc ii} &      -1.43   & 3.89 & 7.87 \\
   311.658 & Fe  {\sc ii}&        -1.50   & 3.89 & 7.87 \\
   311.7666 & Ti  {\sc ii}&       -0.494   & 1.23 & 5.21 \\
   311.8649 & Cr {\sc ii} &        -0.08   & 2.42 & 6.40 \\
   311.9799 & Ti {\sc ii} &        -0.485   & 1.24 & 5.22 \\
   312.0369 & Cr {\sc ii} &        0.10   & 2.43 & 6.41 \\
   312.4978 & Cr {\sc ii} &     0.26  & 2.45 & 6.42 \\
   312.8700 & Cr {\sc ii} &        -0.53  & 2.43 & 6.40 \\
   313.0583 & Be {\sc ii} &              -0.178 & 0.00 & 3.96 \\
   313.1228 & Be {\sc ii} &             -0.479 & 0.00 & 3.96 \\
   313.2056 & Cr {\sc ii} &      0.43   & 2.48 & 6.44 \\
   313.3048 & Fe  {\sc ii}&        -1.9   & 3.89 & 7.84 \\
   313.6681 & Cr {\sc ii} &          -0.44     & 2.45 & 6.41   \\
\hline
\end{tabular}
\end{center}
\end{table}

\subsection {Nova Cir 2018}

 Nova Cir 2018, also PNV J13532700-6725110,    was discovered by John Seach  on  January 19, 2018 in the constellation of Circinus.   The nova brightened  slowly  from  magnitude V $\approx$ 8.5 mag to 6.3 mag by January 27. Then it fluctuated between magnitudes V $\approx$ 6.5 mag and 8.5 mag for about 3 months before fading to below magnitude V $\approx$ 9 mag on June 25th as it can be seen from its light curve shown in Figure \ref{fig:1a}. Spectroscopic observations    obtained
  on 30 January, showed   a number of Fe {\sc ii} lines  with  absorption troughs at  about -1300 \kms \citep{Strader2018}. SALT optical spectroscopy reported a component also at -500 \kms\ \citep{Aydi2018}.  
  
\begin{figure}
\includegraphics[width=0.94\columnwidth]{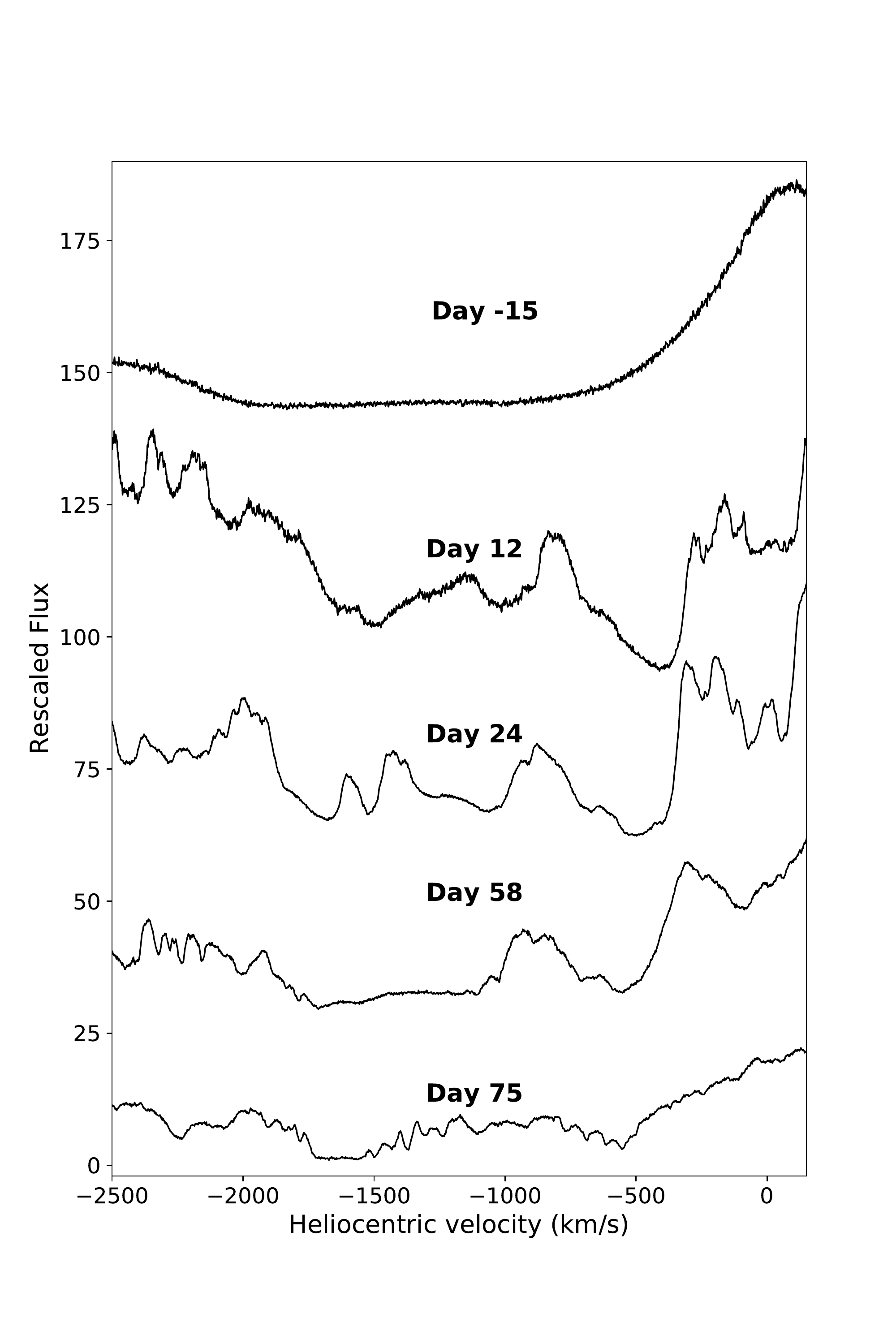}
\caption{  Evolution  of  Nova Cir 2018 in the region of \bevii.  X-axis  as  in Figure \ref{fig:2a}.   }\label{fig:4a}
\end{figure}

The evolution of the spectral features around the  \beviiii $\lambda\lambda$ 313.1 nm region is shown in  Figure \ref{fig:4a}. At day -15 from the maximum taken at JD   2458157.0024  there is  a huge absorption spanning more than one thousand \kms~  that  in the  observation  at day 12     breaks down into two main absorption features  centered  at -400 \kms~ and  -1500 \kms, respectively.  Later   observations   show relatively little change in the profile  with the main absorption feature  shifting to  higher expansion velocities. It is only in the relatively late  spectra  thanks to  the weakening of the  absorption that some fine structure becomes visible. 

\begin{figure*}
\includegraphics[width=0.94\columnwidth]{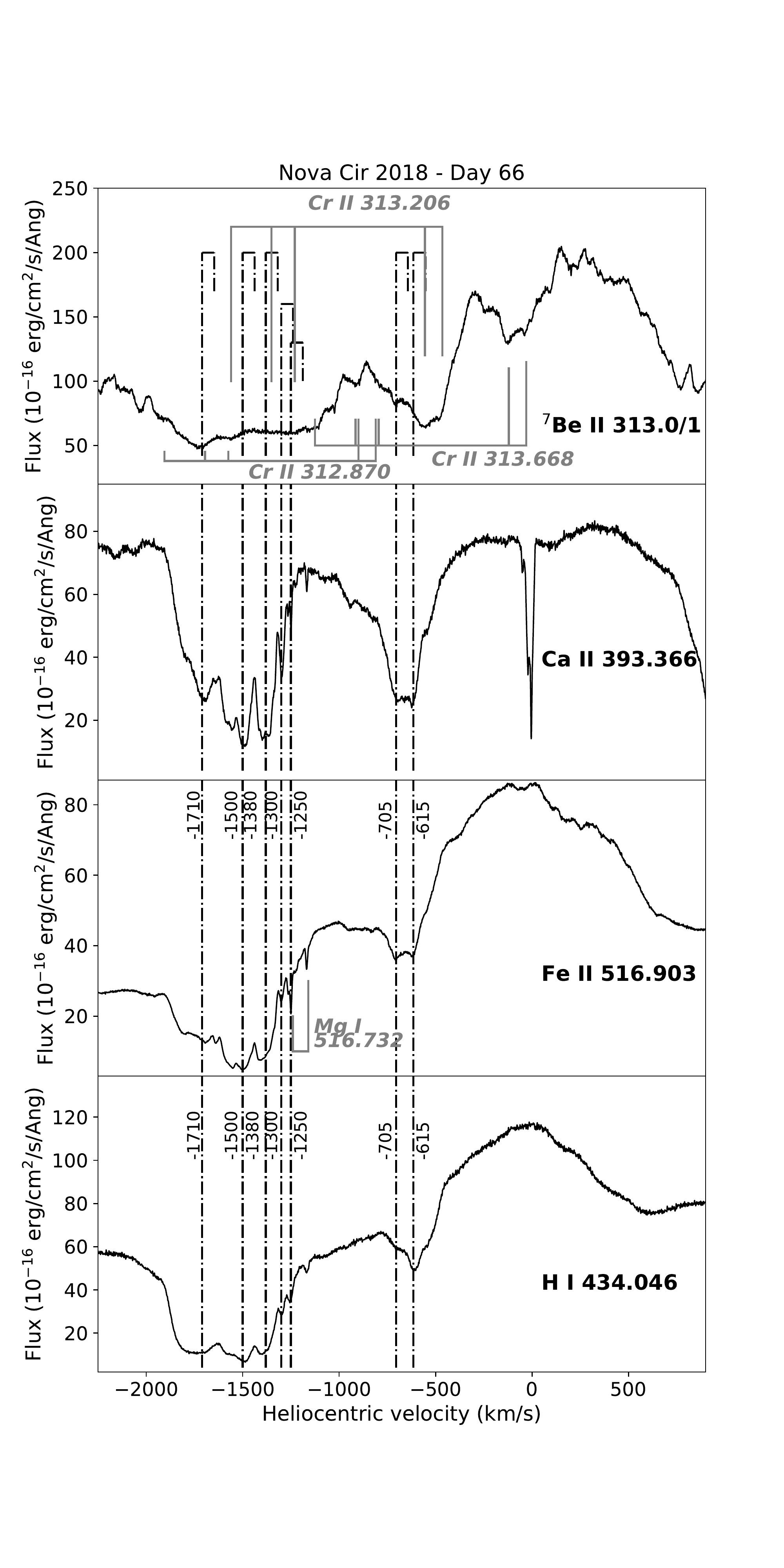}
\includegraphics[width=0.94\columnwidth]{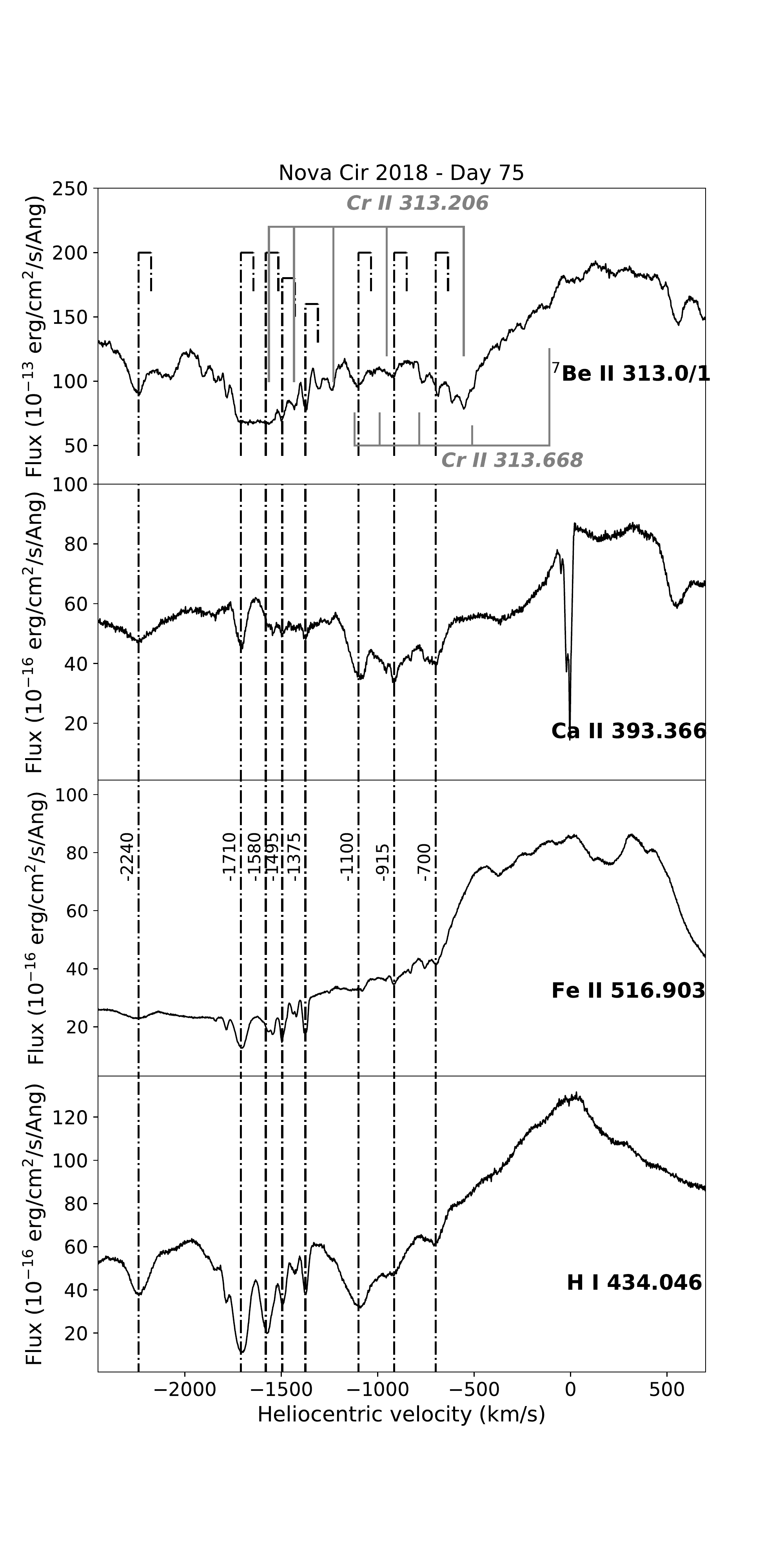}
\caption{  Spectra of   Nova Cir 2018 of day 66 in the left panel  and of day 75  in the right panel.  The \crii\ blends identified in the \beviiii\ panel refer to the five main components with velocities of -615, -705, -1380 , 1500 and -1710 \kms, respectively.}\label{fig:4b}
\end{figure*}

The \beviiii~ region  compared in velocity space  with  other lines  for days 66  and 75 are  shown in   Figure \ref{fig:4b}. At first glance there is some  possible \bevii\ absorption  corresponding in velocity to  the \caii~K features. However, the peak intensity does not correspond precisely and   most of the absorption could be ascribed  to \crii. In particular, there are     narrow features, which became visible only at this stage, without   corresponding   features  of the weaker of the  \beviiii\ doublet. We thus  conclude that  \bevii\  is possibly absent or very weak     in the  outburst spectra of this nova.
  
\subsection {Nova ASASSN-18fv}
ASASSN-18fv is an exceptionally bright nova  discovered in the Carina constellation on March 20,  2018  \citep{Stanek2018},    by the   on going  ASASSN survey. ASASSN-18fv  reached      V $\approx$  6   mag on 22 March. The light curve is characterized by substantial jittering above the base level on time-scale of days.  According to the classification by \citet{Strope2010}, it  was  classified as a J-class nova. ASASSN-18fv was   clearly detected by  NuSTAR in both FPMA and FPMB instruments but not  in the Swift XRT observation (Nelson et al, 2018 ATEL 11608).  
The progenitor of nova ASASSN-18fv
was identified by \citet{strader}
in the Gaia DR1 \citep{Gaia} and 
in the VPHAS DR2 \citep{Drew}. The object in the
Gaia DR1 is at an angular distance of less than 0\farcs{1}
from the nova, with no other objects in the catalogue
within a radius of $5''$. 
The object is also present in Gaia DR2 \citep{DR2,DR2_valid}
that provides a parallax of 0.151 mas with an error of 0.488 mas.
It also provides a G magnitude of 19.688 mag and 
a colour $G_{BP} - G_{RP} = 0.869$ mag.

We succeed to trigger  the   ToO program  on  22 March  achieving  one of the few spectra of a nova during its maximum ever taken.
 A  first report of the    spectroscopic observations   close to the   maximum was  given in  \citet{Izzo2018b}.   The  optical spectrum shows a bright continuum and  is  characterised by  several  narrow absorption features and significant Balmer and Paschen jumps.  The hydrogen lines   the O I 777.3 nm and    several multiplets of \feii\  are in emission   with a P-Cygni profile. The main component in absorption is centered at    v $\sim$ -250 \kms, which  is a relatively modest   velocity for a nova and suggests
 a possible peculiar  nature.  
 
  Our spectra  allow us to obtain an independent
 estimate the reddening in the direction of the nova.
 The reddening maps of \citet{SF11}
provide $E(B-V)=1.1046$ mag for the position of the nova.
In our spectra we were able to detect several interstellar features. The velocity 
span by the Na I D lines and the Ca II H \& K lines
was much smaller than what span by the H I 21\,cm
emission. This strongly suggests that
the light from the nova goes through only a part
of the Galactic interstellar medium along that line of
sight. The Na I D lines are saturated and not well suited to estimate
the reddening, so we used the 
DIB at 578.0\,nm instead. We measure an equivalent
width of 0.01917\, nm that using the relations
of \citet{dib} implies $E(B-V)= 0.37$ mag, i.e. $A_V = 1.15$ mag,
in good agreement with the value from the maps of \citet{SF11}.
Adopting this value and $E(G_{BP}-G_{RP}) =  0.41595 A_V$ \footnote{See \url{http://stev.oapd.inaf.it/cgi-bin/cmd}, this value is derived using the \citet{ODonnell} extinction curve. }  we thus derive a colour for the nova progenitor 
$(G_{BP}-G_{RP})= 0.39066$ mag that corresponds to a black body of
about 8200\,K. 
 
\begin{figure}
\includegraphics[width=0.99\columnwidth]{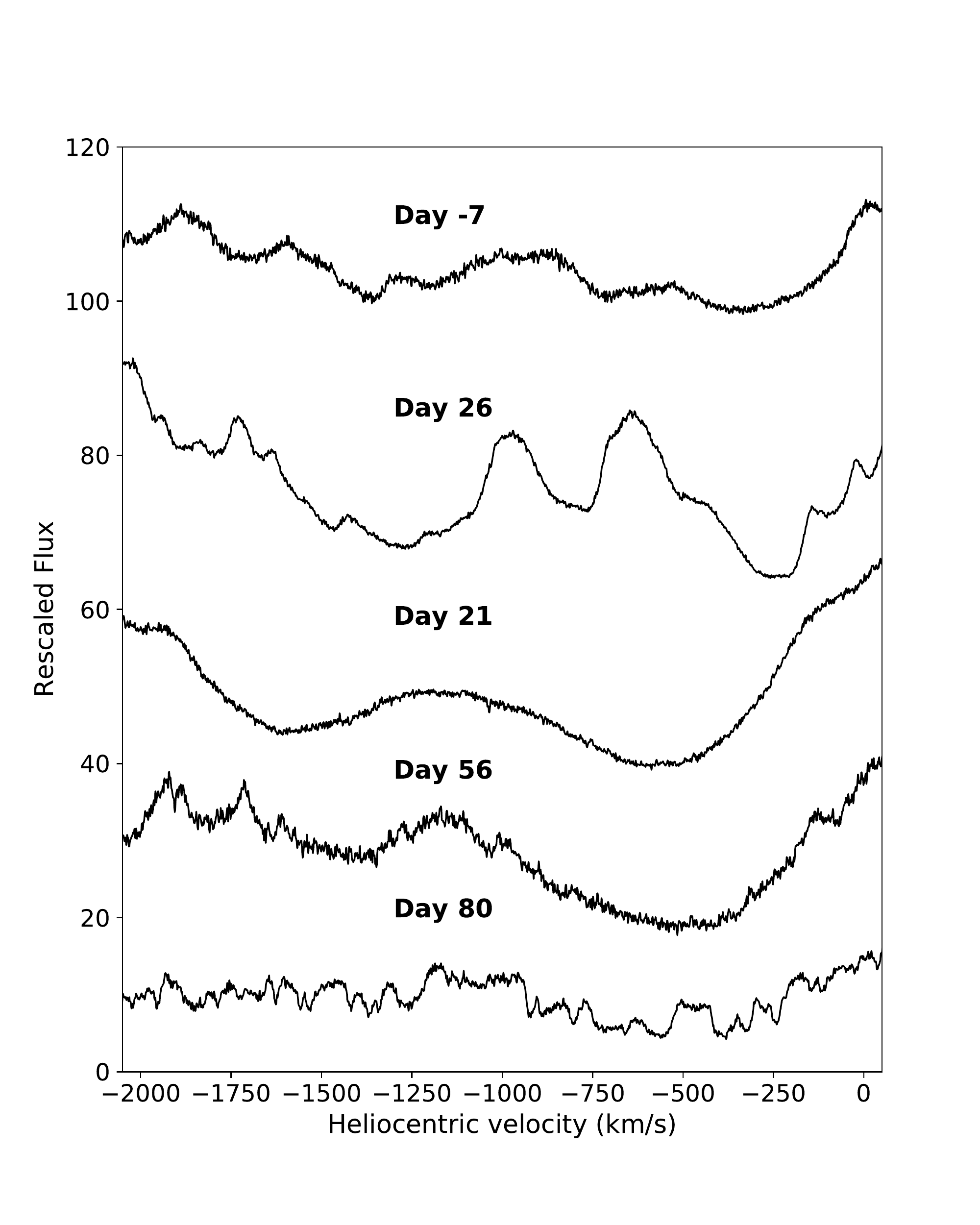}
\caption{  Spectral evolution of   ASASSN-18fv in the region of \bevii. Only a selection of spectra in correspondence of the major spectral changes are shown for sake of clarity. X-axis  as  in Figure \ref{fig:2a}. }\label{fig:5a}
\end{figure}

\begin{figure*}
\includegraphics[width=0.94\columnwidth]{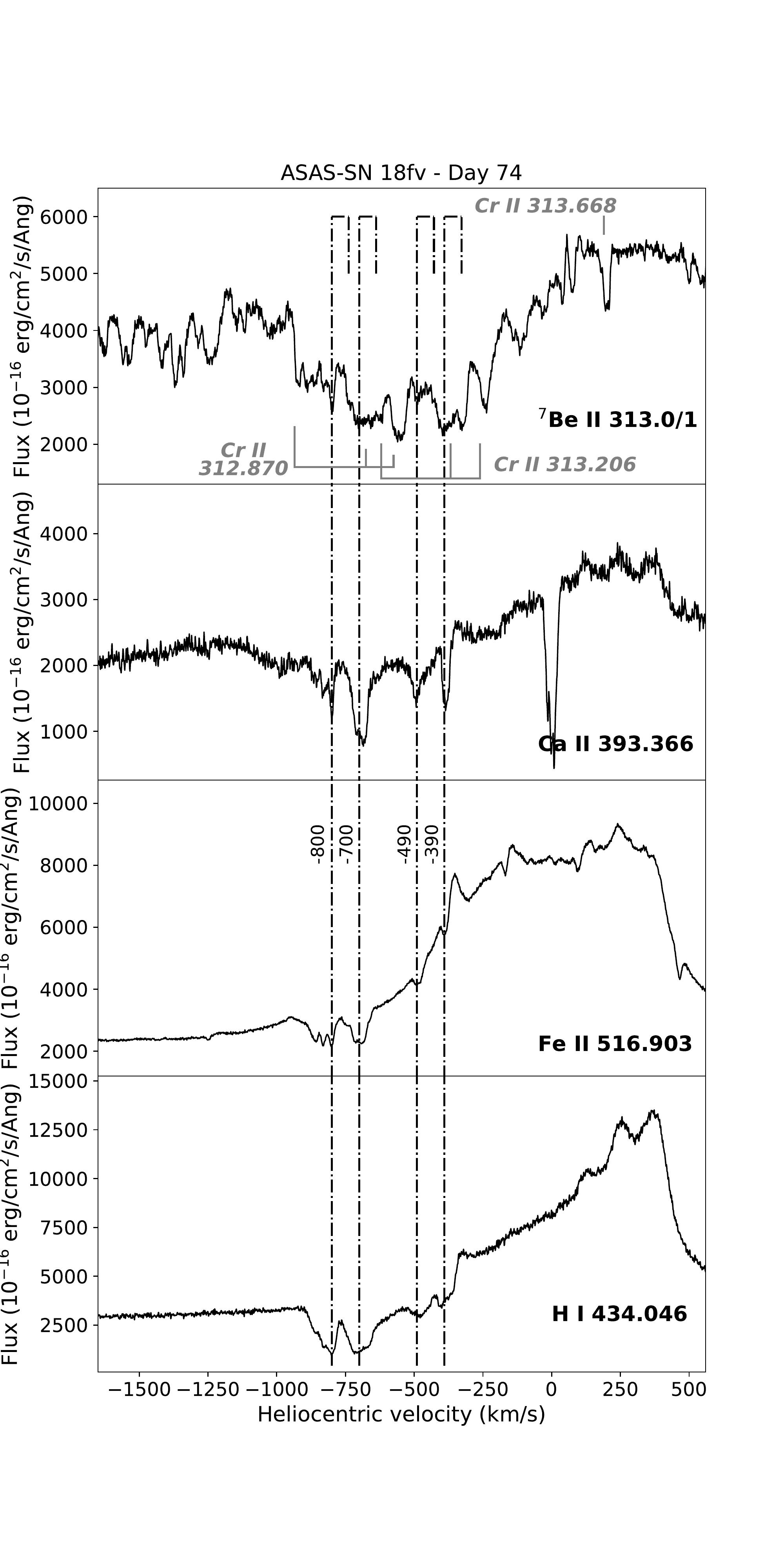}
\includegraphics[width=0.94\columnwidth]{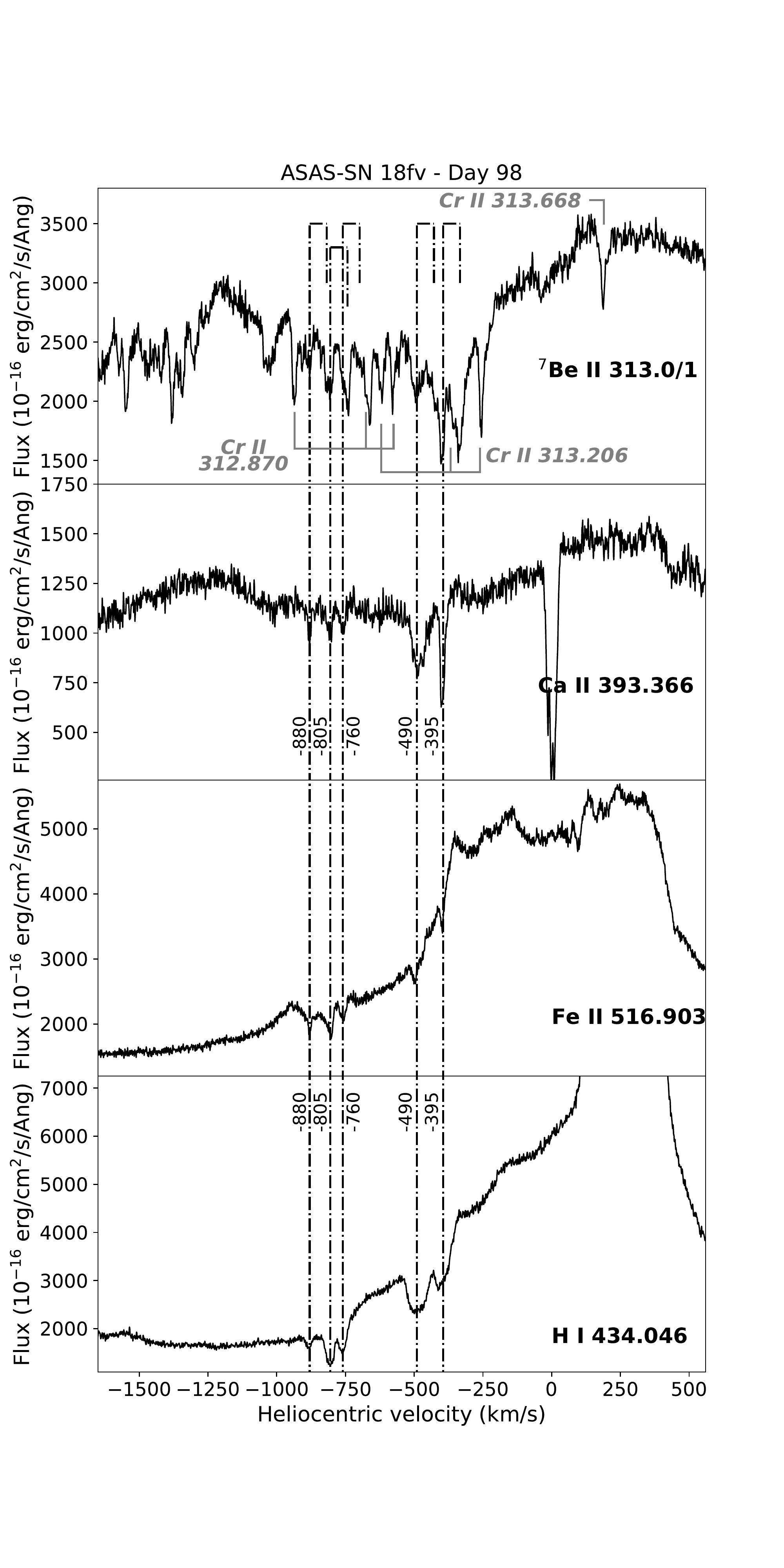}
\caption{  Spectra  of     ASASSN-18fv   at day 98.  The blends identified in the \beviiii\ panel refer to the three main components with velocities of  -390, -490 and -700 \kms, respectively. }\label{fig:5b}
\end{figure*}

\begin{figure}
\includegraphics[width=0.99\columnwidth]{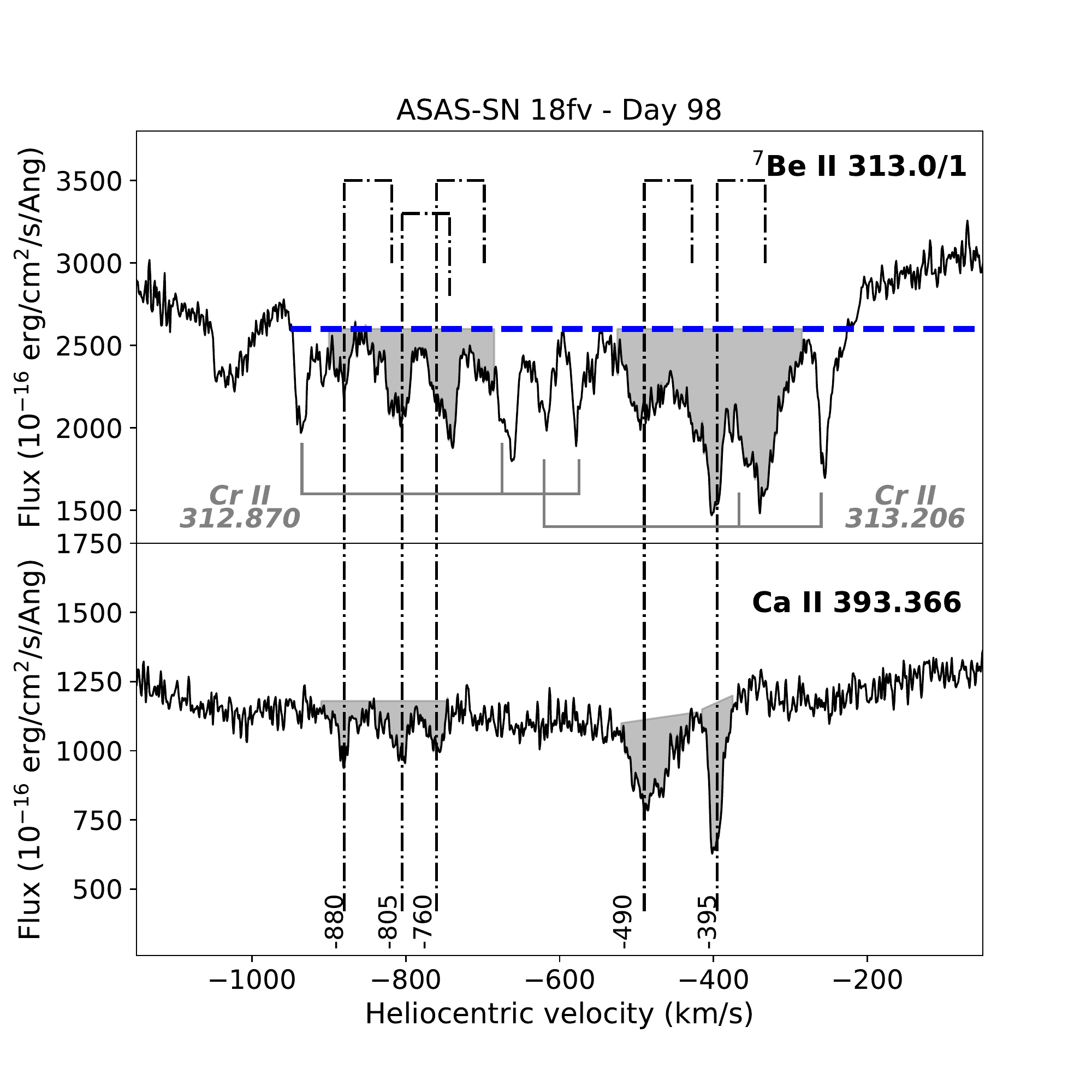}
\caption{   Magnification of the spectrum of day 98 of  ASASSN-18fv. The shadowed regions show the area used for the equivalent widths.}\label{fig:5c}
\end{figure}

In    early epochs  the UVES spectral region around  \bevii~  shows a broad unresolved absorption.  The main contribution is possibly    \beviiii~  but  it cannot be proved. It is only for epochs later than day 63 that the weakening of general absorption reveals structure and the \bevii\ doublet can be recognized in several discrete components {as can be seen in Figure \ref{fig:5a}} . 
 The day 98 spectrum is plotted in Figure \ref{fig:5b}   and  shows  five discrete  components  at -395 , -490, -780, - 805, and -880 \kms, respectively. In correspondence to  these components there is  a feature at the position of  the \beviiii\ $\lambda$ 313.0583 nm line and in several components also  of the  \beviiii\  $\lambda$ 313.1228 nm line.   A magnification  of  the spectrum of this epoch is shown in Figure \ref{fig:5c}.  At this  epoch the \bevii\  decay have reduced by a factor 2 to 3 the original abundance contributing to   the blanketing reduction in the  region. The components of the other lines present in the region as the \crii\ 312.870, 313.206 and 313.668 become  also visible.


\begin{figure}
\includegraphics[width=0.99\columnwidth]{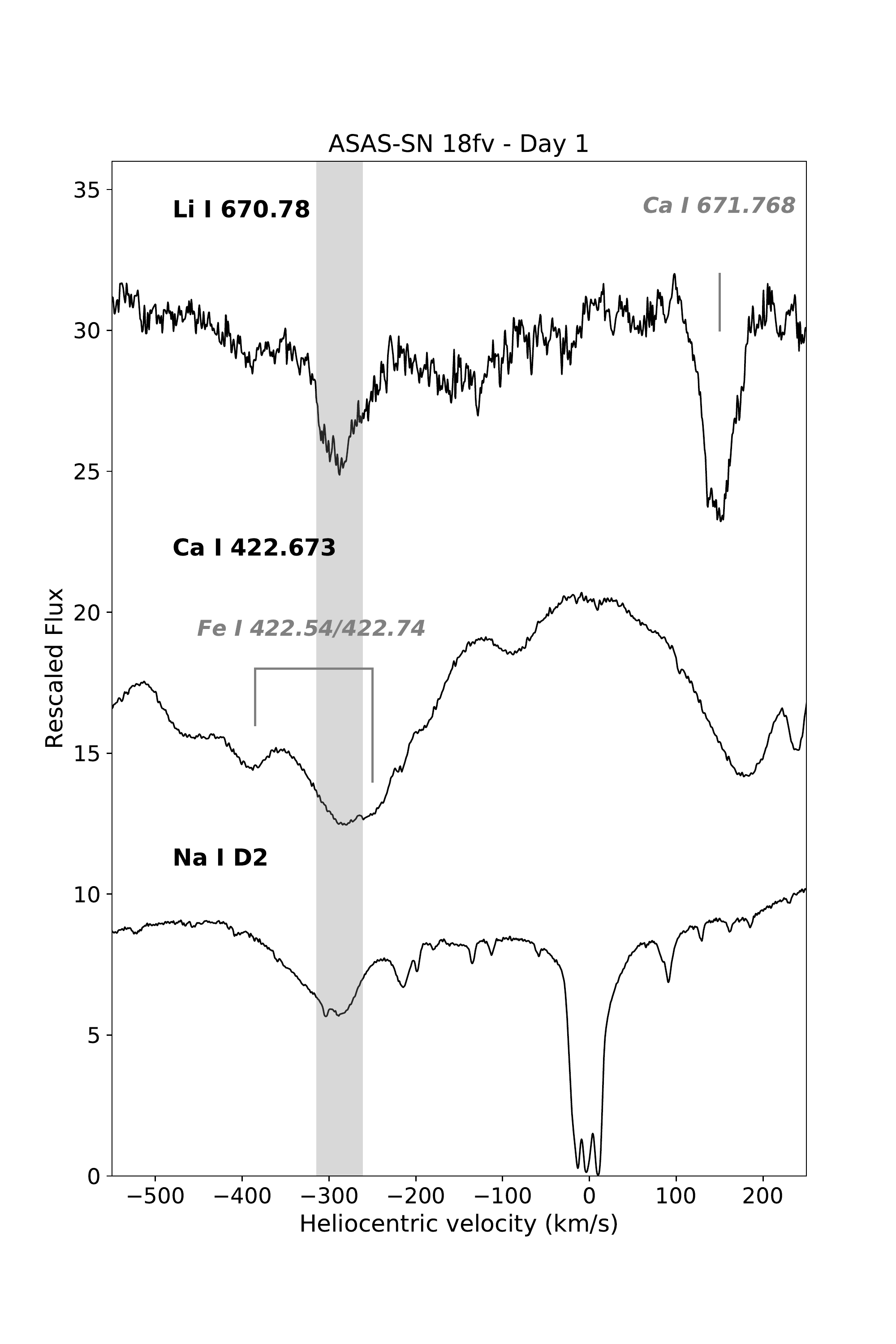}
\caption{   Spectrum of   ASASSN-18fv  at day 1   showing the presence of the absorption feature of \liviii\ $\lambda$670.8 nm, observed  at the same expanding velocity of v = -290 \kms as  the resonance line of \cai\ $\lambda$ 422.673 nm, though  blended with \fei\ $\lambda$ 422.74 nm,  and  the \nai\ D2 absorption. On the top spectrum   the component  at  -290 \kms of the  \cai\ $\lambda$  671.768 nm is also marked. This further support the presence of the neutral species in this component. }\label{fig:5d}
\end{figure}

\section{  \bevii\  Abundance  estimate}

 The  \bevii\ abundance in the nova ejecta can be estimated relatively to Ca,  which is not a nova product.   Singly ionized
 is the main stage for  both species in the expanding shell     with  no evidence of neutral or double ionized stages. However, column densities can be  derived with confidence only when discrete and unsaturated components are seen. These are not always present in the outburst
spectra and sometimes a  sequence of observations lasting about   100 days from maximum are  required for their detection \citep{Molaro2016}.
Moreover, the presence of several blends   and the possibility of emissions  make the  placement of the continuum in the \beviiii~ region  quite problematic. 
To reduce the effects of possible contaminants,  it  is  convenient to measure the sum of the \beviiii\, $\lambda313.0583+\lambda313.1228$  doublet and to compare it with the \caii~  K line at 393.366 nm which is relatively free from blends.   The $\log (gf)$ of  the \beviiii\ doublet are  -0.178, -0.479, and that of 
\caii\,K  is  +0.135. Following  \citet{SpitzerBook} we can write:

\begin{eqnarray}\label{eq:1}
  \frac{N(\mbox{\beviiii})}{N(\mbox{\caii})}
  & = &   2.164   \times \frac{W(\mbox{\beviiii\,, Doublet})}{W(\mbox{\caii\,, K})} 
\end{eqnarray}

 The Nova Mus 2018 spectrum in the regions of   \beviiii~ and \caii\ K line at day 45 is shown in  Figure \ref{fig:3c}.  A major contaminant is  \crii\    and the  expected positions of the strongest lines of  Table \ref{tab:3} are  marked in the figure. The component at -820 \kms of  \crii~ 313.668 nm is  identified. The same component of the \crii~ 313.206    is  strong  and has  an EW of 96 m\AA. The  components at -920 and -1150 \kms  should be present but  blended with   the \beviiii~ absorption. Also marked in the figure are the components of the \crii~ 312.870 nm line. However,  the -1150 \kms component should be  visible if present and, therefore,     this    line is   very weak. The \crii 312.498 nm and the \crii~  312.036  lines are   seen in all the main velocity components. By scaling with the relative strength  we  estimate that  the equivalent widths of   \crii\ components blended within  the \beviiii\ absorption  are of  160  m\AA\  from   \crii\ 313.206 and of $\approx$  40 m\AA\  from  \crii\ 312.870 nm.
The equivalent width  of the whole region spanned by the \beviiii\ components  is of 1.8 $\pm $ 0.150 \AA\ with blends  accounted for. The
main  uncertainty  comes from  the continuum placement with a possible error of about 10\%. On the other hand, the \caii\ 393.3 nm  line is rather free of blends and the local continuum can be also determined with confidence. The equivalent width  of the \caii\ 393.3 nm  line is of
0.910 $ \pm 0.030 $ \AA\  providing  a ratio of   EW(\beviiii)/EW(\caii) = 1.98. Note that by using  the component at velocity -1150 \kms alone  and  the stronger of the \beviiiii\ doublet  we would get a ratio of  EW(\beviiii$_{313.0583} $  / EW(\caii) = 1.22. This  is equivalent to a ratio of 1.83  for a  whole \bevii\ absorption and consistent with the ratio derived by using the integral of the whole absorption including all components.
With  Eq. \ref{eq:1}  we obtain   N(\beviiii)/N(\caii) = 4.28 ($\pm$ 0.4). \bevii\  is unstable with    a  half life of 53.2  days. Thus, the    abundance at  explosion  has been reduced by a factor of 1.6 after 35 days.  Assuming a  solar abundance  of  \caii/H = 2.18 $\cdot 10^{-6}$ \citep{Lodders2009},  we obtain in Nova Mus 2018  an abundance of   \bevii/H = 1.5 $\cdot 10^{-5}$. 
The Calcium abundance is assumed to be solar in all cases here considered.  Should Ca  be higher than  solar,   the final \bevii /H would decrease  accordingly.

 The ASASSN-18fv UVES spectrum  at day 98  is shown in Figure \ref{fig:5b}.  There are  five discrete and narrow  \caii\ K  components  and all have    the corresponding  \beviiii\ 313.0583 nm and  313.1228 nm lines.  At this epoch also the   contaminants can be seen and    Cr\,{\sc ii}\    312.8700 nm, 
Cr\,{\sc ii}\  313.2053 nm    and  Cr\,{\sc ii}\  313.6681 nm are identified in the Figure.
The ratio  between the shadowed area  of Be II and Ca II  in Figure \ref{fig:5c} is EW(\bevii)/EW(\caii) =  1.52. Similar ratios are measured also at  day 63, 81  and 86. Thus, N( \beviiii) / N(\caii)  = 3.29, by number.  Considering the \bevii\ decay the original abundance becomes  a factor  3.0 larger, i.e.  N(\beviiii)/N(\caii) = 9.9.  Assuming also here  that all of Ca and \bevii\ are singly ionized  and solar abundances for Calcium,   we obtain in ASASSN-18fv an abundance of  \bevii/H = 2.15 x $10^{-5}$.
 
\section{Discussion}

 The present detection of \bevii~ in the outburst spectra of ASASSN-18fv and Nova Mus 2018   provides additional  evidence that \bevii\  is freshly created in the
thermonuclear runaway via the reaction  \iiihe($\alpha,\gamma$)\bevii\  and ejected during  nova explosion.  Nova Mus 2018 is a fast nova while ASASSN-18fv is  moderately fast, according to the \citet{Payne1957} classification, showing that the \bevii\ production is not related to the kind of nova.  This is consistent   with what observed in the sample of nova studied so far with   a mixture of different nova types and  uncorrelated  \bevii\ abundances.

In   ASASSN-17hx and   possibly also in  Nova Cir  2018 there is no clear evidence for  the presence of \bevii\ in the outburst spectrum.    These are the first novae  where the \bevii\ isotope is not detected since  it  has been searched for. This   shows  that the thermonuclear reaction chain taking place onto the WD and the ejection phase could be  quite complex.
The \bevii\ produced in the thermonuclear runaway chain needs to be transported relatively quickly to the surface by convection. This is  a sort of  Cameron-Fowler mechanism  invoked for Li-rich red giants \citep{CameronFowler1971} and  may not always operate  effectively.  The turbulent turnover timescales  are of the order of 100 sec throughout the  initial deflagration stage. Therefore, at the same time as the 
temperature is driving the convection the mixing may be  
incomplete within the shell  \citep{Shore2019}.
 We note also that  in the latest numerical simulations of \citet{starrfield2019} the \bevii\ yields are very sensitive to the WD mass, decreasing by a factor of 30 from the most massive WD to the lighter ones.

The extant \bevii/H measurements are summarized in Table \ref{tab:3}. The abundances derived  in Nova Mus 2018  and ADSASSN18xv are    comparable to those derived in  V339 Del  \citep{Tajitsu2015},  V838 Her \citep{Selvelli2018} and V2944 Oph  \citep{Tajitsu2016} but a factor 3 lower  than those obtained in V407 Lup   \citep{Izzo2018} and a factor 5 lower  than  in V5668 Sgr  \citep{Molaro2016}.  
 
 We emphasize that the  \bevii/H  remain  larger   by at least one order of magnitude than predicted by nova  models. \citep{Starrfield1978,Hernanz1996,Jose1998}. 
We note, however,    that the final amount of \bevii~  is  sensitive to the amount of \iiihe\ in the donor star   since a higher abundance of \iiihe\ is expected to result in a higher \bevii\ abundance.
\citet{Boffin1993} and \citet{Hernanz1996} found a logarithmic dependence of the \bevii\ output to the initial \iiihe\ abundance.
The non linearity of the \bevii\ yields results from   \iiihe(\iiihe, 2p)\ivhe\
which  importance increases as the square of the initial \iiihe\ abundance  and therefore producing a  leaking of the available \iiihe\ for the \iiihe ($\alpha, \gamma $) \bevii\  which rate increases only linearly of the initial \iiihe\ abundance.
For \iiihe\ enhancements  up to one hundred solar  
\citet{Boffin1993} derive
$X($\bevii$)/X($\bevii$_0) = 1 + 1.5\log X($\iiihe$)/X($\iiihe$_{\odot})$,
where  \bevii$_0 $ is the \bevii\ final mass fraction obtained with a solar initial \iiihe\ mass fraction. 
From the theoretical point of view it is  believed that low mass  main sequence stars     synthesizes \iiihe\ through  the p-p chains with peak abundances of few $10^{-3}$ by number \citep{Iben1967}. As the star ascends the red giant branch convection dredge up  \iiihe\ enriched material to the surface which is later expelled into the interstellar medium by wind or during the planetary phase.
\iiihe\ is a particularly  difficult element to measure. It can be measured in H~II regions by  using measurements at a frequency of 8.665 GHz, i.e. 3.46 cm, which is emitted naturally by ionized $^3$He \citep{Bania2010,Balser2018} or in   stellar atmospheres of hot stars \citep{Geier2012}. Surprisingly, interstellar medium  observations indicate that there is far less of this element in the Galaxy than the current models predict.  In order not to overproduce  \iiihe\  in the course of chemical evolution, it has become customary to assume that some unknown \iiihe-destruction mechanism is at work in  low-mass giants \citep{Dearborn1996,Galli1997,Chiappini2002,Romano2003}. For instance,  \citet{Charbonnel2007}    suggested a  thermohaline mixing  during the red giant branch phase of low-mass stars. 
However, in a few planetary nebulae  \iiihe/H  is found high at the level of  10 $^{-3}$ quite  consistent with predictions from standard stellar models \citep{Rood1992,Balser2018}.

Novae are semi-detached  binary systems in which a Roche lobe filling secondary
star transfers material to the WD  primary. In
the Roche  geometry there is a direct relation between the orbital period  and the  mass of the
secondary star  (Warner 1995, \citep{Warner1976,Knigge2006}),  that, in the case of
novae, gives M2 values close to  0.3 $M_{\odot}$.   Since stars with mass  lower  than  0.3 $M_{\odot}$
 are fully convective   the \iiihe\  produced in the core by incomplete pp1 chain is transported outward and transferred  to the WD
surface.  
If the donor star has  \iiihe\ much greater than the solar value this would almost reconcile  the observations with the model predictions.


 \bevii\    decays via electron-capture into   \livii\ ~    with a half-life of 53.22 days. However, despite    the  temporal span of our observations, that  in some cases extend up to more than  a hundred days, we do not detect    the \liviii\, $\lambda\lambda$670.8 nm  absorption line  at the corresponding positions  of other neutral lines such as    Na\,{\sc i}\ D doublet  and Ca\,{\sc i}\,  $\lambda$422.7 nm  which are observed.  We do not detect any  \liviii\, $\lambda\lambda$670.8 nm  absorption in correspondence of the main components of Nova Mus 2018.   A possible  weak feature is detected in the very early spectrum of  Nova ASASSN-18fv as shown in Figure \ref{fig:5d}. So far,  the \liviii\ 670.8 nm line  has   been detected only
 in the early spectra  of  Nova Cen 2013  {\bf   \citep{Izzo2015}.}
 This shows   that the physical conditions in the ejecta    sometimes permit   the survival of  neutral  \liviii\  at least in the very early stages. However, this cannot be the result of \bevii\ decay since it is close to the nova outburst. As argued by \citet{Molaro2016} the  \bevii\    decays  via capture  of an internal K electron resulting  into   \liviiii\ ~    which has no  lines in the optical spectral range 
 and therefore is not observable. The non detection of \livii\   in the later  outburst spectra   implies  that the ejected gas had been heated and   that almost all  Li   remains ionised. 

The typical atomic fraction \bevii /H of  $\approx$ 2 $\times$ 10 $^{-5}$ corresponds to 
a \livii ~  overabundance of   four  orders of magnitude    with respect to the  meteoritic value of \livii /H $\approx$ 2 $\times$ 10$^{-9}$\citep{Lodders2009}. With an ejecta of $\approx$   10$^{-5}$ $M_{\odot}$ a nova event is producing a \livii\ mass of $1.4 \times$ 10$^{-9}$ $M_{\odot}^{Li}$ and, with a nova rate of 50 novae yr$^{-1}$ along a  10$^{10}$ yrs, a total \livii\  mass of 700 $M_{\odot}^{Li} $, which is  a considerable fraction of the about 1000 $M_{\odot}^{Li} $ estimate for the Galaxy \citep{Cescutti2019}.
The precise fraction contributed by  novae is related to uncertainties in the  nova rate and to the behaviour of the rate in the course of Galactic life. However, the primordial \livii\ production is between the  $\approx$ 80 $M_{\odot}^{Li} $ when taking the halo stars  abundance,  and 250 $M_{\odot}^{Li} $, when taking  the predicted value of primordial nucleosynthesis   with the baryon density of the deuterium abundance or CMB.  The  contribution by galactic cosmic rays is of about   8 $M_{\odot}^{Li} $ as  estimated indirectly by  $^9$Be  which is produced by the same processes \citep{Molaro1997}. The  contribution of the Asymptotic Giant Branch stars  is  about  1 $M_{\odot}^{Li} $ and rather negligible \citep{Rukeya2017}.     The nova \livii\ production  has been considered by means of a detailed model of the chemical evolution of the Milky Way  by  \citet{Cescutti2019}. They    showed that novae could   account  for the observed increase of \livii\  abundances with increasing of metallicity in the thin disk. The agreement of the model with the Li abundances is obtained for a delay time for the nova production of $\approx$ 1 Gyr and of $1.8 (\pm 0.6) \times 10^{-5}$ M$_{\odot}^{Li}$ of \livii~ as effective yields for a whole  nova life, which is consistent with the mean \bevii\ observed. With $\approx 10^{4}$ nova events during a  lifetime this correspond to a $1.8  \times 10^{-9}$ M$_{\odot}^{Li}$ per event, which is consistent with what obtained here. Despite all uncertainties novae   appear as the dominant source and the only one able to account for the bulk of Galactic \livii\ .

\begin{table}
\caption{The   \bevii/H~ (number)  for the  Novae with narrow
absorption components. The original values from \citet{Tajitsu2015,Tajitsu2016} are corrected here for the
decay of \bevii. References are: (1) \citet{Tajitsu2015}; (2)
\citet{Molaro2016},  (3) \citet{Izzo2018}, (4) \citet{Tajitsu2016} (5) \citet{Selvelli2018} and (6) this paper.}  \label{tab:3}
\begin{center}
\begin{tabular}{llrrllrr}
\hline
\hline

\multicolumn{1}{c}{Nova} &
\multicolumn{1}{c}{type} &
\multicolumn{1}{c}{day} &
\multicolumn{1}{c}{comp} &
\multicolumn{1}{c}{ {\bevii/H~} } &
\multicolumn{1}{c}{Ref} \\
&&&&&\\
\hline
\hline

V339 Del & CO & 47 & -1103& 1.9 $\times$ 10 $^{-5}$  & 1\\

 &  &47  &-1268  &3.2 $\times$ 10 $^{-5}$   & 1 \\

V5668 Sgr & CO &  58 & -1175 & 1.7 $\times$ 10 $^{-4}$  & 2 \\
 &  & 82 &-1500 &1.3 $\times$ 10 $^{-4}$  & 2 \\
 V2944 Oph  & CO &  80 &  -645 & 1.6 $\times$ 10 $^{-5}$  & 4 \\
V407 Lup  & ONe&  8  &-2030 & 6.2  $\times$ 10 $^{-5}$ & 3 \\
V838 Her & &    &  &   2 $\times$ 10 $^{-5}$ & 5 \\
ASASSN-17hx & &    &  &   none & 6 \\
Nova Mus & &  35  & all &   1.5$\times$ 10 $^{-5}$ & 6 \\
Nova Cir & &    & &    uncertain & 6 \\
ASASSN-18fv & &  80  & all &   2.2$\times$ 10 $^{-5}$ & 6\\
\hline
\end{tabular}
\end{center}

\end{table}

\section{Conclusions }

Following the recent detection  of \beviiii ~  in the   outburst spectra of  Classical Novae  we started a ToO project  at ESO with the high resolution spectrograph UVES to  search for the \bevii~ isotope in all bright novae. The  nova brightness  is required  to study  the  resonance doublet lines \beviiii ~ at $\lambda\lambda$313.0583,\,313.1228 nm  where the atmospheric absorption is strong and optical elements of spectrograph have low efficiency. 
We observed all the four bright novae of the last two years. We summarize here the main results.

\begin{itemize}
    \item{The \beviiii~ doublet  absorption lines    are   detected in two of them, Nova  Mus 2018 and ASASSN-18fv, confirming   the synthesis of \bevii~  in the nova  thermonuclear runaway. 
The      atomic fractions  by number  are estimated to be  $X(\mbox{\bevii})/X(\mbox{H})$   $\approx$ 1.5 $\cdot 10 ^{-5}$ and 2.2 $\cdot 10 ^{-5}$ in Nova Mus 2018 and ASASSN-18fv, respectively, when  the   \bevii\ decay is taken into account.
There are 7 novae where the \bevii\ abundance has been measured. Five of them have an \bevii/H abundance close to  $\approx$  2 $\cdot 10 ^{-5}$ while two show higher values. The value of $\approx$  2 $\cdot 10 ^{-5}$ looks like  a typical abundance, though the sample  remain rather small.}

\item{ We do not detect \beviiii\ lines
in the spectra of  ASASSN-17hx and  possibly neither in Nova Cir 2018. This shows that not all novae eject  \bevii.  Theoretically it is expected 
that the \bevii\ yields decrease with the decreasing of the WD mass. The yields decrease by a factor of 45 passing form  1.35 to 0.6 M$_{\odot}$ WD \citep{starrfield2019}. Moreover,  very little \bevii\ is produced in the case of reduced  mixing between the WD core products with the material of the donor star  and the absence of \bevii\ could reveal such an occurrence. The present fraction of novae without evidence of \bevii\ is of 22\%.  To note that  \citet{Mason2019} recently suggested that ASASSN-17hx  could not be  a classical nova or at   least a very peculiar one.  If this is the case, the absence of \bevii\ would have a very specific meaning.}    

\item{The \bevii/H abundance of   $\approx$  2 $\cdot 10 ^{-5}$  is higher by about one order of magnitude than the predictions of  nova  models. We argue that an  higher  than solar  abundance of \iiihe\ in the donor star would  result in higher \bevii\ yields. In fact \iiihe /H $\approx 10^{-3} $, one hundred times the solar  as observed in few planetary nebulae by \citet{Balser2018}, could be common in the small mass donor stars thus  producing a factor four increase in the \bevii\ yields. This would   reduce significantly  the disagreement between   the observations and the nova models. }

\item{ A  \bevii/H~ abundance of  $\approx$  2 $\cdot 10 ^{-5}$    implies  a  \livii/H    overproduction of    $\approx$   4 dex  above  the meteoritic value. A simple estimation  based on the mass ejecta and nova rate shows that they    are  likely the missing   \livii~  source required to account for  the   75-90\% of the   \livii\    in  the Milky Way. }

\section*{Acknowledgments}
We gratefully acknowledge Elena Mason and Gabriele Cescutti for  helpful discussions on
 the interpretation of the data. We thank an anonymous  referee for many useful comments.  We also  thank   Gabriella Schiulaz for checking the English language. LI was supported by grants from VILLUM FONDEN (project numbers 16599 and 25501)

 \end{itemize}


\begin{thebibliography}{}
\makeatletter
\relax
\def\mn@urlcharsother{\let\do\@makeother \do\$\do\&\do\#\do\^\do\_\do\%\do\~}
\def\mn@doi{\begingroup\mn@urlcharsother \@ifnextchar [ {\mn@doi@}
  {\mn@doi@[]}}
\def\mn@doi@[#1]#2{\def\@tempa{#1}\ifx\@tempa\@empty \href
  {http://dx.doi.org/#2} {doi:#2}\else \href {http://dx.doi.org/#2} {#1}\fi
  \endgroup}
\def\mn@eprint#1#2{\mn@eprint@#1:#2::\@nil}
\def\mn@eprint@arXiv#1{\href {http://arxiv.org/abs/#1} {{\tt arXiv:#1}}}
\def\mn@eprint@dblp#1{\href {http://dblp.uni-trier.de/rec/bibtex/#1.xml}
  {dblp:#1}}
\def\mn@eprint@#1:#2:#3:#4\@nil{\def\@tempa {#1}\def\@tempb {#2}\def\@tempc
  {#3}\ifx \@tempc \@empty \let \@tempc \@tempb \let \@tempb \@tempa \fi \ifx
  \@tempb \@empty \def\@tempb {arXiv}\fi \@ifundefined
  {mn@eprint@\@tempb}{\@tempb:\@tempc}{\expandafter \expandafter \csname
  mn@eprint@\@tempb\endcsname \expandafter{\@tempc}}}

\bibitem[\protect\citeauthoryear{{Arenou} et~al.,}{{Arenou}
  et~al.}{2018}]{DR2_valid}
{Arenou} F.,  et~al., 2018, \mn@doi [\aap] {10.1051/0004-6361/201833234}, \href
  {https://ui.adsabs.harvard.edu/abs/2018A&A...616A..17A} {616, A17}

\bibitem[\protect\citeauthoryear{{Arnould} \& {Norgaard}}{{Arnould} \&
  {Norgaard}}{1975}]{Arnould1975}
{Arnould} M.,  {Norgaard} H.,  1975, \aap, \href
  {http://adsabs.harvard.edu/abs/1975A26A....42...55A} {42, 55}

\bibitem[\protect\citeauthoryear{{Aydi}, {Buckley}, {Mohamed}  \&
  {Whitelock}}{{Aydi} et~al.}{2018}]{Aydi2018}
{Aydi} E.,  {Buckley} D.~A.~H.,  {Mohamed} S.,   {Whitelock} P.~A.,  2018, The
  Astronomer's Telegram, \href
  {https://ui.adsabs.harvard.edu/abs/2018ATel11287....1A} {11287, 1}

\bibitem[\protect\citeauthoryear{{Balser} \& {Bania}}{{Balser} \&
  {Bania}}{2018}]{Balser2018}
{Balser} D.~S.,  {Bania} T.~M.,  2018, \mn@doi [\aj]
  {10.3847/1538-3881/aaeb2b}, \href
  {https://ui.adsabs.harvard.edu/abs/2018AJ....156..280B} {156, 280}

\bibitem[\protect\citeauthoryear{{Bania}, {Rood}  \& {Balser}}{{Bania}
  et~al.}{2010}]{Bania2010}
{Bania} T.~M.,  {Rood} R.~T.,   {Balser} D.~S.,  2010, in {Charbonnel} C.,
  {Tosi} M.,  {Primas} F.,   {Chiappini} C.,  eds,  IAU Symposium Vol. 268,
  Light Elements in the Universe. pp 81--90, \mn@doi{10.1017/S174392131000390X}

\bibitem[\protect\citeauthoryear{{Boffin}, {Paulus}, {Arnould}  \&
  {Mowlavi}}{{Boffin} et~al.}{1993}]{Boffin1993}
{Boffin} H.~M.~J.,  {Paulus} G.,  {Arnould} M.,   {Mowlavi} N.,  1993, \aap,
  \href {http://adsabs.harvard.edu/abs/1993A%26A...279..173B} {279, 173}

\bibitem[\protect\citeauthoryear{{Bonifacio}, {Monai}  \& {Beers}}{{Bonifacio}
  et~al.}{2000}]{BMB}
{Bonifacio} P.,  {Monai} S.,   {Beers} T.~C.,  2000, \mn@doi [\aj]
  {10.1086/301566}, \href
  {https://ui.adsabs.harvard.edu/abs/2000AJ....120.2065B} {120, 2065}

\bibitem[\protect\citeauthoryear{{Cameron} \& {Fowler}}{{Cameron} \&
  {Fowler}}{1971}]{CameronFowler1971}
{Cameron} A.~G.~W.,  {Fowler} W.~A.,  1971, \mn@doi [\apj] {10.1086/150821},
  \href {http://adsabs.harvard.edu/abs/1971ApJ...164..111C} {164, 111}

\bibitem[\protect\citeauthoryear{{Cescutti} \& {Molaro}}{{Cescutti} \&
  {Molaro}}{2019}]{Cescutti2019}
{Cescutti} G.,  {Molaro} P.,  2019, \mn@doi [\mnras] {10.1093/mnras/sty2967},
  \href {http://adsabs.harvard.edu/abs/2019MNRAS.482.4372C} {482, 4372}

\bibitem[\protect\citeauthoryear{{Charbonnel} \& {Zahn}}{{Charbonnel} \&
  {Zahn}}{2007}]{Charbonnel2007}
{Charbonnel} C.,  {Zahn} J.-P.,  2007, \mn@doi [\aap]
  {10.1051/0004-6361:20077274}, \href
  {https://ui.adsabs.harvard.edu/abs/2007A%26A...467L..15C} {467, L15}

\bibitem[\protect\citeauthoryear{{Chiappini}, {Renda}  \&
  {Matteucci}}{{Chiappini} et~al.}{2002}]{Chiappini2002}
{Chiappini} C.,  {Renda} A.,   {Matteucci} F.,  2002, \mn@doi [\aap]
  {10.1051/0004-6361:20021314}, \href
  {https://ui.adsabs.harvard.edu/abs/2002A&A...395..789C} {395, 789}

\bibitem[\protect\citeauthoryear{{Dearborn}, {Steigman}  \& {Tosi}}{{Dearborn}
  et~al.}{1996}]{Dearborn1996}
{Dearborn} D.~S.~P.,  {Steigman} G.,   {Tosi} M.,  1996, \mn@doi [\apj]
  {10.1086/177472}, \href
  {https://ui.adsabs.harvard.edu/abs/1996ApJ...465..887D} {465, 887}

\bibitem[\protect\citeauthoryear{{Drew} et~al.,}{{Drew} et~al.}{2014}]{Drew}
{Drew} J.~E.,  et~al., 2014, \mn@doi [\mnras] {10.1093/mnras/stu394}, \href
  {https://ui.adsabs.harvard.edu/abs/2014MNRAS.440.2036D} {440, 2036}

\bibitem[\protect\citeauthoryear{{Friedjung}}{{Friedjung}}{1979}]{Friedjung1979}
{Friedjung} M.,  1979, \aap, \href
  {http://adsabs.harvard.edu/abs/1979A\%26A....77..357F} {77, 357}

\bibitem[\protect\citeauthoryear{{Friedman} et~al.,}{{Friedman}
  et~al.}{2011}]{dib}
{Friedman} S.~D.,  et~al., 2011, \mn@doi [\apj] {10.1088/0004-637X/727/1/33},
  \href {https://ui.adsabs.harvard.edu/abs/2011ApJ...727...33F} {727, 33}

\bibitem[\protect\citeauthoryear{{Gaia Collaboration} et~al.,}{{Gaia
  Collaboration} et~al.}{2016}]{Gaia}
{Gaia Collaboration} et~al., 2016, \mn@doi [\aap]
  {10.1051/0004-6361/201629272}, \href
  {https://ui.adsabs.harvard.edu/abs/2016A&A...595A...1G} {595, A1}

\bibitem[\protect\citeauthoryear{{Gaia Collaboration} et~al.,}{{Gaia
  Collaboration} et~al.}{2018}]{DR2}
{Gaia Collaboration} et~al., 2018, \mn@doi [\aap]
  {10.1051/0004-6361/201833051}, \href
  {https://ui.adsabs.harvard.edu/abs/2018A&A...616A...1G} {616, A1}

\bibitem[\protect\citeauthoryear{{Galli}, {Stanghellini}, {Tosi}  \&
  {Palla}}{{Galli} et~al.}{1997}]{Galli1997}
{Galli} D.,  {Stanghellini} L.,  {Tosi} M.,   {Palla} F.,  1997, \mn@doi [\apj]
  {10.1086/303708}, \href
  {https://ui.adsabs.harvard.edu/abs/1997ApJ...477..218G} {477, 218}

\bibitem[\protect\citeauthoryear{{Gaposchkin}}{{Gaposchkin}}{1957}]{Payne1957}
{Gaposchkin} C.~H.~P.,  1957, {The galactic novae.}

\bibitem[\protect\citeauthoryear{{Geier}, {Heber}, {Edelmann}, {Morales-Rueda},
  {Kilkenny}, {O'Donoghue}, {Marsh}  \& {Copperwheat}}{{Geier}
  et~al.}{2012}]{Geier2012}
{Geier} S.,  {Heber} U.,  {Edelmann} H.,  {Morales-Rueda} L.,  {Kilkenny} D.,
  {O'Donoghue} D.,  {Marsh} T.~R.,   {Copperwheat} C.,  2012, in {Kilkenny} D.,
   {Jeffery} C.~S.,   {Koen} C.,  eds,  Astronomical Society of the Pacific
  Conference Series Vol. 452, Fifth Meeting on Hot Subdwarf Stars and Related
  Objects. p.~57 (\mn@eprint {arXiv} {1112.2918})

\bibitem[\protect\citeauthoryear{{Gomez-Gomar}, {Hernanz}, {Jose}  \&
  {Isern}}{{Gomez-Gomar} et~al.}{1998}]{Gomez1998}
{Gomez-Gomar} J.,  {Hernanz} M.,  {Jose} J.,   {Isern} J.,  1998, \mn@doi
  [\mnras] {10.1046/j.1365-8711.1998.01421.x}, \href
  {https://ui.adsabs.harvard.edu/abs/1998MNRAS.296..913G} {296, 913}

\bibitem[\protect\citeauthoryear{{Guarro} et~al.,}{{Guarro}
  et~al.}{2017}]{Guarro2017}
{Guarro} J.,  et~al., 2017, The Astronomer's Telegram, \href
  {https://ui.adsabs.harvard.edu/abs/2017ATel10737....1G} {10737, 1}

\bibitem[\protect\citeauthoryear{{Harris}, {Teegarden}, {Weidenspointner},
  {Palmer}, {Cline}, {Gehrels}  \& {Ramaty}}{{Harris}
  et~al.}{2001}]{Harris2001}
{Harris} M.~J.,  {Teegarden} B.~J.,  {Weidenspointner} G.,  {Palmer} D.~M.,
  {Cline} T.~L.,  {Gehrels} N.,   {Ramaty} R.,  2001, \mn@doi [\apj]
  {10.1086/323951}, \href
  {https://ui.adsabs.harvard.edu/abs/2001ApJ...563..950H} {563, 950}

\bibitem[\protect\citeauthoryear{{Hernanz}, {Jose}, {Coc}  \&
  {Isern}}{{Hernanz} et~al.}{1996}]{Hernanz1996}
{Hernanz} M.,  {Jose} J.,  {Coc} A.,   {Isern} J.,  1996, \mn@doi [\apjl]
  {10.1086/310122}, \href {http://adsabs.harvard.edu/abs/1996ApJ...465L..27H}
  {465, L27}

\bibitem[\protect\citeauthoryear{{Iben}}{{Iben}}{1967}]{Iben1967}
{Iben} Icko J.,  1967, \mn@doi [\araa] {10.1146/annurev.aa.05.090167.003035},
  \href {https://ui.adsabs.harvard.edu/abs/1967ARA&A...5..571I} {5, 571}

\bibitem[\protect\citeauthoryear{{Izzo} et~al.,}{{Izzo}
  et~al.}{2015}]{Izzo2015}
{Izzo} L.,  et~al., 2015, \mn@doi [\apjl] {10.1088/2041-8205/808/1/L14}, \href
  {http://adsabs.harvard.edu/abs/2015ApJ...808L..14I} {808, L14}

\bibitem[\protect\citeauthoryear{{Izzo} et~al.,}{{Izzo}
  et~al.}{2018a}]{Izzo2018}
{Izzo} L.,  et~al., 2018a, \mn@doi [\mnras] {10.1093/mnras/sty435}, \href
  {http://adsabs.harvard.edu/abs/2018MNRAS.478.1601I} {478, 1601}

\bibitem[\protect\citeauthoryear{{Izzo} et~al.,}{{Izzo}
  et~al.}{2018b}]{Izzo2018b}
{Izzo} L.,  et~al., 2018b, The Astronomer's Telegram, \href
  {http://adsabs.harvard.edu/abs/2018ATel11468....1I} {11468}

\bibitem[\protect\citeauthoryear{{Jos{\'e}} \& {Hernanz}}{{Jos{\'e}} \&
  {Hernanz}}{1998}]{Jose1998}
{Jos{\'e}} J.,  {Hernanz} M.,  1998, \mn@doi [\apj] {10.1086/305244}, \href
  {http://adsabs.harvard.edu/abs/1998ApJ...494..680J} {494, 680}

\bibitem[\protect\citeauthoryear{{Knigge}}{{Knigge}}{2006}]{Knigge2006}
{Knigge} C.,  2006, \mn@doi [\mnras] {10.1111/j.1365-2966.2006.11096.x}, \href
  {https://ui.adsabs.harvard.edu/abs/2006MNRAS.373..484K} {373, 484}

\bibitem[\protect\citeauthoryear{{Lawler}, {Sneden}, {Nave}, {Den Hartog},
  {Emraho{\u{g}}lu}  \& {Cowan}}{{Lawler} et~al.}{2017}]{lawler2017L}
{Lawler} J.~E.,  {Sneden} C.,  {Nave} G.,  {Den Hartog} E.~A.,
  {Emraho{\u{g}}lu} N.,   {Cowan} J.~J.,  2017, \mn@doi [\apjs]
  {10.3847/1538-4365/228/1/10}, \href
  {https://ui.adsabs.harvard.edu/abs/2017ApJS..228...10L} {228, 10}

\bibitem[\protect\citeauthoryear{{Lodders}, {Palme}  \& {Gail}}{{Lodders}
  et~al.}{2009}]{Lodders2009}
{Lodders} K.,  {Palme} H.,   {Gail} H.-P.,  2009, \mn@doi [Landolt
  B{\"o}rnstein] {10.1007/978-3-540-88055-4_34}, \href
  {http://cdsads.u-strasbg.fr/abs/2009LanB...4B...44L} {}

\bibitem[\protect\citeauthoryear{{Mason}, {Shore}, {Kuin}  \&
  {Bohlsen}}{{Mason} et~al.}{2019}]{Mason2019}
{Mason} E.,  {Shore} S.,  {Kuin} P.,   {Bohlsen} T.,  2019, \aap \, submitted

\bibitem[\protect\citeauthoryear{{Molaro}, {Bonifacio}, {Castelli}  \&
  {Pasquini}}{{Molaro} et~al.}{1997}]{Molaro1997}
{Molaro} P.,  {Bonifacio} P.,  {Castelli} F.,   {Pasquini} L.,  1997, \aap,
  \href {https://ui.adsabs.harvard.edu/abs/1997A&A...319..593M} {319, 593}

\bibitem[\protect\citeauthoryear{{Molaro}, {Izzo}, {Mason}, {Bonifacio}  \&
  {Della Valle}}{{Molaro} et~al.}{2016}]{Molaro2016}
{Molaro} P.,  {Izzo} L.,  {Mason} E.,  {Bonifacio} P.,   {Della Valle} M.,
  2016, \mn@doi [\mnras] {10.1093/mnrasl/slw169}, \href
  {http://adsabs.harvard.edu/abs/2016MNRAS.463L.117M} {463, L117}

\bibitem[\protect\citeauthoryear{{Munari}, {Ochner}, {Hambsch}, {Frigo},
  {Castellani}, {Milani}, {Valisa}  \& {Vagnozzi}}{{Munari}
  et~al.}{2017}]{Munari2017}
{Munari} U.,  {Ochner} P.,  {Hambsch} F.-J.,  {Frigo} A.,  {Castellani} F.,
  {Milani} A.,  {Valisa} P.,   {Vagnozzi} A.,  2017, The Astronomer's Telegram,
  \href {https://ui.adsabs.harvard.edu/abs/2017ATel10736....1M} {10736}

\bibitem[\protect\citeauthoryear{{O'Donnell}}{{O'Donnell}}{1994}]{ODonnell}
{O'Donnell} J.~E.,  1994, \mn@doi [\apj] {10.1086/173713}, \href
  {https://ui.adsabs.harvard.edu/abs/1994ApJ...422..158O} {422, 158}

\bibitem[\protect\citeauthoryear{{Pavana}, {Anupama}, {Selvakumar}  \&
  {Kiran}}{{Pavana} et~al.}{2017}]{Pavana2017}
{Pavana} M.,  {Anupama} G.~C.,  {Selvakumar} G.,   {Kiran} B.~S.,  2017, The
  Astronomer's Telegram, \href
  {https://ui.adsabs.harvard.edu/abs/2017ATel10613....1P} {10613, 1}

\bibitem[\protect\citeauthoryear{{Poggiani}}{{Poggiani}}{2018}]{Poggiani2018}
{Poggiani} R.,  2018, arXiv e-prints, \href
  {https://ui.adsabs.harvard.edu/abs/2018arXiv180707947P} {p. arXiv:1807.07947}

\bibitem[\protect\citeauthoryear{{Romano} \& {Matteucci}}{{Romano} \&
  {Matteucci}}{2003}]{Romano2003}
{Romano} D.,  {Matteucci} F.,  2003, \mn@doi [\mnras]
  {10.1046/j.1365-8711.2003.06526.x}, \href
  {https://ui.adsabs.harvard.edu/abs/2003MNRAS.342..185R} {342, 185}

\bibitem[\protect\citeauthoryear{{Rood}, {Bania}  \& {Wilson}}{{Rood}
  et~al.}{1992}]{Rood1992}
{Rood} R.~T.,  {Bania} T.~M.,   {Wilson} T.~L.,  1992, \mn@doi [\nat]
  {10.1038/355618a0}, \href
  {https://ui.adsabs.harvard.edu/abs/1992Natur.355..618R} {355, 618}

\bibitem[\protect\citeauthoryear{Rukeya, Lü, Wang  \& Zhu}{Rukeya
  et~al.}{2017}]{Rukeya2017}
Rukeya R.,  Lü G.,  Wang Z.,   Zhu C.,  2017, \mn@doi [Publications of the
  Astronomical Society of the Pacific] {10.1088/1538-3873/aa6b4d}, 129, 074201

\bibitem[\protect\citeauthoryear{{Schlafly} \& {Finkbeiner}}{{Schlafly} \&
  {Finkbeiner}}{2011}]{SF11}
{Schlafly} E.~F.,  {Finkbeiner} D.~P.,  2011, \mn@doi [\apj]
  {10.1088/0004-637X/737/2/103}, \href
  {https://ui.adsabs.harvard.edu/abs/2011ApJ...737..103S} {737, 103}

\bibitem[\protect\citeauthoryear{{Schlegel}, {Finkbeiner}  \&
  {Davis}}{{Schlegel} et~al.}{1998}]{Schlegel}
{Schlegel} D.~J.,  {Finkbeiner} D.~P.,   {Davis} M.,  1998, \mn@doi [\apj]
  {10.1086/305772}, \href
  {https://ui.adsabs.harvard.edu/abs/1998ApJ...500..525S} {500, 525}

\bibitem[\protect\citeauthoryear{{Selvelli}, {Molaro}  \& {Izzo}}{{Selvelli}
  et~al.}{2018}]{Selvelli2018}
{Selvelli} P.,  {Molaro} P.,   {Izzo} L.,  2018, \mn@doi [\mnras]
  {10.1093/mnras/sty2310}, \href
  {http://adsabs.harvard.edu/abs/2018MNRAS.481.2261S} {481, 2261}

\bibitem[\protect\citeauthoryear{{Shappee} et~al.,}{{Shappee}
  et~al.}{2014}]{Shappee2014}
{Shappee} B.~J.,  et~al., 2014, \mn@doi [\apj] {10.1088/0004-637X/788/1/48},
  \href {http://adsabs.harvard.edu/abs/2014ApJ...788...48S} {788, 48}

\bibitem[\protect\citeauthoryear{{Shore}}{{Shore}}{2019}]{Shore2019}
{Shore} S.~N.,  2019, in {Werner} K.,  {Stehle} C.,  {Rauch} T.,   {Lanz} T.,
  eds,  Astronomical Society of the Pacific Conference Series Vol. 519,
  Astronomical Society of the Pacific Conference Series. p.~161

\bibitem[\protect\citeauthoryear{{Siegert} et~al.,}{{Siegert}
  et~al.}{2018}]{Siegert2018}
{Siegert} T.,  et~al., 2018, \mn@doi [\aap] {10.1051/0004-6361/201732514},
  \href {http://adsabs.harvard.edu/abs/2018A%26A...615A.107S} {615, A107}

\bibitem[\protect\citeauthoryear{{Spitzer}}{{Spitzer}}{1998}]{SpitzerBook}
{Spitzer} L.,  1998, {Physical Processes in the Interstellar Medium}

\bibitem[\protect\citeauthoryear{{Stanek} et~al.,}{{Stanek}
  et~al.}{2017a}]{Stanek2017c}
{Stanek} K.~Z.,  et~al., 2017a, The Astronomer's Telegram, \href
  {http://adsabs.harvard.edu/abs/2017ATel10387....1S} {10387}

\bibitem[\protect\citeauthoryear{{Stanek} et~al.,}{{Stanek}
  et~al.}{2017b}]{Stanek2017b}
{Stanek} K.~Z.,  et~al., 2017b, The Astronomer's Telegram, \href
  {http://adsabs.harvard.edu/abs/2017ATel10436....1S} {10436}

\bibitem[\protect\citeauthoryear{{Stanek} et~al.,}{{Stanek}
  et~al.}{2017c}]{Stanek2017a}
{Stanek} K.~Z.,  et~al., 2017c, The Astronomer's Telegram, \href
  {http://adsabs.harvard.edu/abs/2017ATel10523....1S} {10523}

\bibitem[\protect\citeauthoryear{{Stanek} et~al.,}{{Stanek}
  et~al.}{2018}]{Stanek2018}
{Stanek} K.~Z.,  et~al., 2018, The Astronomer's Telegram, \href
  {http://adsabs.harvard.edu/abs/2018ATel11454....1S} {11454}

\bibitem[\protect\citeauthoryear{{Starrfield}, {Truran}, {Sparks}  \&
  {Arnould}}{{Starrfield} et~al.}{1978}]{Starrfield1978}
{Starrfield} S.,  {Truran} J.~W.,  {Sparks} W.~M.,   {Arnould} M.,  1978,
  \mn@doi [\apj] {10.1086/156175}, \href
  {http://adsabs.harvard.edu/abs/1978ApJ...222..600S} {222, 600}

\bibitem[\protect\citeauthoryear{{Starrfield}, {Iliadis}  \&
  {Hix}}{{Starrfield} et~al.}{2016}]{Starrfield2016}
{Starrfield} S.,  {Iliadis} C.,   {Hix} W.~R.,  2016, \mn@doi [\pasp]
  {10.1088/1538-3873/128/963/051001}, \href
  {http://adsabs.harvard.edu/abs/2016PASP..128e1001S} {128, 051001}

\bibitem[\protect\citeauthoryear{{Starrfield}, {Bose}, {Iliadis}, {Hix},
  {Woodward}  \& {Wagner}}{{Starrfield} et~al.}{2019}]{starrfield2019}
{Starrfield} S.,  {Bose} M.,  {Iliadis} C.,  {Hix} W.~R.,  {Woodward} C.~E.,
  {Wagner} R.~M.,  2019, arXiv e-prints, \href
  {https://ui.adsabs.harvard.edu/abs/2019arXiv191000575S} {p. arXiv:1910.00575}

\bibitem[\protect\citeauthoryear{{Strader}, {Chomiuk}, {Swihart}  \&
  {Shishkovsky}}{{Strader} et~al.}{2018a}]{Strader2018}
{Strader} J.,  {Chomiuk} L.,  {Swihart} S.,   {Shishkovsky} L.,  2018a, The
  Astronomer's Telegram, \href
  {https://ui.adsabs.harvard.edu/abs/2018ATel11209....1S} {11209, 1}

\bibitem[\protect\citeauthoryear{{Strader}, {Chomiuk}, {Holoien}, {Prieto},
  {Stanek}, {Shappee}  \& {Dong}}{{Strader} et~al.}{2018b}]{strader}
{Strader} J.,  {Chomiuk} L.,  {Holoien} T.~W.~S.,  {Prieto} J.~L.,  {Stanek}
  K.~Z.,  {Shappee} B.~J.,   {Dong} S.,  2018b, The Astronomer's Telegram,
  \href {https://ui.adsabs.harvard.edu/abs/2018ATel11456....1S} {11456, 1}

\bibitem[\protect\citeauthoryear{{Strope}, {Schaefer}  \& {Henden}}{{Strope}
  et~al.}{2010}]{Strope2010}
{Strope} R.~J.,  {Schaefer} B.~E.,   {Henden} A.~A.,  2010, \mn@doi [\aj]
  {10.1088/0004-6256/140/1/34}, \href
  {https://ui.adsabs.harvard.edu/abs/2010AJ....140...34S} {140, 34}

\bibitem[\protect\citeauthoryear{{Tajitsu}, {Sadakane}, {Naito}, {Arai}  \&
  {Aoki}}{{Tajitsu} et~al.}{2015}]{Tajitsu2015}
{Tajitsu} A.,  {Sadakane} K.,  {Naito} H.,  {Arai} A.,   {Aoki} W.,  2015,
  \mn@doi [\nat] {10.1038/nature14161}, \href
  {http://adsabs.harvard.edu/abs/2015Natur.518..381T} {518, 381}

\bibitem[\protect\citeauthoryear{{Tajitsu}, {Sadakane}, {Naito}, {Arai},
  {Kawakita}  \& {Aoki}}{{Tajitsu} et~al.}{2016}]{Tajitsu2016}
{Tajitsu} A.,  {Sadakane} K.,  {Naito} H.,  {Arai} A.,  {Kawakita} H.,   {Aoki}
  W.,  2016, \mn@doi [\apj] {10.3847/0004-637X/818/2/191}, \href
  {http://adsabs.harvard.edu/abs/2016ApJ...818..191T} {818, 191}

\bibitem[\protect\citeauthoryear{{Warner}}{{Warner}}{1976}]{Warner1976}
{Warner} B.,  1976, in {Eggleton} P.,  {Mitton} S.,   {Whelan} J.,  eds,  IAU
  Symposium Vol. 73, Structure and Evolution of Close Binary Systems. p.~85

\bibitem[\protect\citeauthoryear{{Williams} \& {Darnley}}{{Williams} \&
  {Darnley}}{2017}]{Williams2017}
{Williams} S.~C.,  {Darnley} M.~J.,  2017, The Astronomer's Telegram, \href
  {https://ui.adsabs.harvard.edu/abs/2017ATel10542....1W} {10542, 1}

\makeatother
\end{thebibliography}
\end{document}